\documentclass[twocolumn]{aastex62}

\newcommand{\allsne}{5,243}
\newcommand{\unsupsne}{2,885}
\newcommand{\specsne}{557}

\newcommand{\newsne}{2,315} 

\newcommand{\numslsne}{17}
\newcommand{\numiip}{94}
\newcommand{\numiin}{24}
\newcommand{\numia}{404}
\newcommand{\numibc}{19}

\newcommand{\newslsne}{37}
\newcommand{\newiip}{459}
\newcommand{\newiin}{112}
\newcommand{\newia}{1435}
\newcommand{\newibc}{272}

\newcommand{\allslsne}{54}
\newcommand{\alliip}{553}
\newcommand{\alliin}{136}
\newcommand{\allia}{1839}
\newcommand{\allibc}{291}

\newcommand{\allslsneper}{1.6}
\newcommand{\alliipper}{19.8}
\newcommand{\alliinper}{4.8}
\newcommand{\alliaper}{62.0}
\newcommand{\allibcper}{11.7}

\usepackage{amsmath}

\graphicspath{{./}{figures/}}

\received{January 1, 2018}
\revised{January 7, 2018}
\accepted{\today}
\submitjournal{ApJ}

\shorttitle{SuperRAENN}
\shortauthors{Villar et al.}

\begin{document}
\newcommand{\LCO}{\affiliation{Las Cumbres Observatory, 6740 Cortona Drive, Suite 102, Goleta, CA 93117-5575, USA}}
\newcommand{\Simons}{\affiliation{Simons Junior Fellow, Department of Astronomy, Columbia University, New York, NY 10027-6601, USA}}
\newcommand{\DSI}{\affiliation{Harvard Data Science Initiative, Harvard University, Cambridge, MA 02138, USA}}
\newcommand{\berkman}{\affiliation{Berkman Klein Center, Harvard University, Cambridge, MA 02138}}
\newcommand{\UCSB}{\affiliation{Department of Physics, University of California, Santa Barbara, CA 93106-9530, USA}}
\newcommand{\KITP}{\affiliation{Kavli Institute for Theoretical Physics, University of California, Santa Barbara, CA 93106-4030, USA}}
\newcommand{\UCD}{\affiliation{Department of Physics, University of California, 1 Shields Avenue, Davis, CA 95616-5270, USA}}
\newcommand{\WIS}{\affiliation{Department of Particle Physics and Astrophysics, Weizmann Institute of Science, 76100 Rehovot, Israel}}
\newcommand{\OKC}{\affiliation{Oskar Klein Centre, Department of Astronomy, Stockholm University, Albanova University Centre, SE-106 91 Stockholm, Sweden}}
\newcommand{\OAPD}{\affiliation{INAF-Osservatorio Astronomico di Padova, Vicolo dell'Osservatorio 5, I-35122 Padova, Italy}}
\newcommand{\Caltech}{\affiliation{Cahill Center for Astronomy and Astrophysics, California Institute of Technology, Mail Code 249-17, Pasadena, CA 91125, USA}}
\newcommand{\GSFC}{\affiliation{Astrophysics Science Division, NASA Goddard Space Flight Center, Mail Code 661, Greenbelt, MD 20771, USA}}
\newcommand{\UMD}{\affiliation{Joint Space-Science Institute, University of Maryland, College Park, MD 20742, USA}}
\newcommand{\UCB}{\affiliation{Department of Astronomy, University of California, Berkeley, CA 94720-3411, USA}}
\newcommand{\TTU}{\affiliation{Department of Physics, Texas Tech University, Box 41051, Lubbock, TX 79409-1051, USA}}
\newcommand{\STScI}{\affiliation{Space Telescope Science Institute, 3700 San Martin Drive, Baltimore, MD 21218-2410, USA}}
\newcommand{\UT}{\affiliation{University of Texas at Austin, 1 University Station C1400, Austin, TX 78712-0259, USA}}
\newcommand{\IoA}{\affiliation{Institute of Astronomy, University of Cambridge, Madingley Road, Cambridge CB3 0HA, UK}}
\newcommand{\QUB}{\affiliation{Astrophysics Research Centre, School of Mathematics and Physics, Queen's University Belfast, Belfast BT7 1NN, UK}}
\newcommand{\IPAC}{\affiliation{Spitzer Science Center, California Institute of Technology, Pasadena, CA 91125, USA}}
\newcommand{\JPL}{\affiliation{Jet Propulsion Laboratory, California Institute of Technology, 4800 Oak Grove Dr, Pasadena, CA 91109, USA}}
\newcommand{\Southampton}{\affiliation{Department of Physics and Astronomy, University of Southampton, Southampton SO17 1BJ, UK}}
\newcommand{\LANL}{\affiliation{Space and Remote Sensing, MS B244, Los Alamos National Laboratory, Los Alamos, NM 87545, USA}}
\newcommand{\Tsinghua}{\affiliation{Physics Department and Tsinghua Center for Astrophysics, Tsinghua University, Beijing, 100084, People's Republic of China}}
\newcommand{\NAOC}{\affiliation{National Astronomical Observatory of China, Chinese Academy of Sciences, Beijing, 100012, People's Republic of China}}
\newcommand{\Itagaki}{\affiliation{Itagaki Astronomical Observatory, Yamagata 990-2492, Japan}}
\newcommand{\Einstein}{\altaffiliation{Einstein Fellow}}
\newcommand{\Hubble}{\altaffiliation{Hubble Fellow}}
\newcommand{\CfA}{\affiliation{Center for Astrophysics \textbar{} Harvard \& Smithsonian, 60 Garden Street, Cambridge, MA 02138-1516, USA}}
\newcommand{\UA}{\affiliation{Department of Astronomy/Steward Observatory, 933 North Cherry Avenue, Tucson, AZ 85721-0065, USA}}
\newcommand{\MPA}{\affiliation{Max-Planck-Institut f\"ur Astrophysik, Karl-Schwarzschild-Stra\ss e 1, D-85748 Garching, Germany}}
\newcommand{\DSFP}{\altaffiliation{LSSTC Data Science Fellow}}
\newcommand{\HCO}{\affiliation{Harvard College Observatory, 60 Garden Street, Cambridge, MA 02138-1516, USA}}
\newcommand{\Carnegie}{\affiliation{Observatories of the Carnegie Institute for Science, 813 Santa Barbara Street, Pasadena, CA 91101-1232, USA}}
\newcommand{\TAU}{\affiliation{School of Physics and Astronomy, Tel Aviv University, Tel Aviv 69978, Israel}}
\newcommand{\Edinburgh}{\affiliation{Institute for Astronomy, University of Edinburgh, Royal Observatory, Blackford Hill EH9 3HJ, UK}}
\newcommand{\Birmingham}{\affiliation{Birmingham Institute for Gravitational Wave Astronomy and School of Physics and Astronomy, University of Birmingham, Birmingham B15 2TT, UK}}
\newcommand{\CIERA}{\affiliation{Center for Interdisciplinary Exploration and Research in Astrophysics and Department of Physics and Astronomy, \\Northwestern University, 2145 Sheridan Road, Evanston, IL 60208-3112, USA}}
\newcommand{\Bath}{\affiliation{Department of Physics, University of Bath, Claverton Down, Bath BA2 7AY, UK}}
\newcommand{\CTIO}{\affiliation{Cerro Tololo Inter-American Observatory, National Optical Astronomy Observatory, Casilla 603, La Serena, Chile}}
\newcommand{\Potsdam}{\affiliation{Institut f\"ur Physik und Astronomie, Universit\"at Potsdam, Haus 28, Karl-Liebknecht-Str. 24/25, D-14476 Potsdam-Golm, Germany}}
\newcommand{\INPE}{\affiliation{Instituto Nacional de Pesquisas Espaciais, Avenida dos Astronautas 1758, 12227-010, S\~ao Jos\'e dos Campos -- SP, Brazil}}
\newcommand{\UNC}{\affiliation{Department of Physics and Astronomy, University of North Carolina, 120 East Cameron Avenue, Chapel Hill, NC 27599, USA}}
\newcommand{\Ohio}{\affiliation{Astrophysical Institute, Department of Physics and Astronomy, 251B Clippinger Lab, Ohio University, Athens, OH 45701-2942, USA}}
\newcommand{\AAS}{\affiliation{American Astronomical Society, 1667 K~Street NW, Suite 800, Washington, DC 20006-1681, USA}}
\newcommand{\MMT}{\affiliation{MMT and Steward Observatories, University of Arizona, 933 North Cherry Avenue, Tucson, AZ 85721-0065, USA}}
\newcommand{\Geneva}{\affiliation{ISDC, Department of Astronomy, University of Geneva, Chemin d'\'Ecogia, 16 CH-1290 Versoix, Switzerland}}
\newcommand{\IUCAA}{\affiliation{Inter-University Center for Astronomy and Astrophysics, Post Bag 4, Ganeshkhind, Pune, Maharashtra 411007, India}}
\newcommand{\CMU}{\affiliation{Department of Physics, Carnegie Mellon University, 5000 Forbes Avenue, Pittsburgh, PA 15213-3815, USA}}
\newcommand{\NAOJ}{\affiliation{Division of Science, National Astronomical Observatory of Japan, 2-21-1 Osawa, Mitaka, Tokyo 181-8588, Japan}}
\newcommand{\IfA}{\affiliation{Institute for Astronomy, University of Hawai`i, 2680 Woodlawn Drive, Honolulu, HI 96822-1839, USA}}
\newcommand{\UCSC}{\affiliation{Department of Astronomy and Astrophysics, University of California, Santa Cruz, CA 95064-1077, USA}}
\newcommand{\Purdue}{\affiliation{Department of Physics and Astronomy, Purdue University, 525 Northwestern Avenue, West Lafayette, IN 47907-2036, USA}}
\newcommand{\Princeton}{\affiliation{Department of Astrophysical Sciences, Princeton University, 4 Ivy Lane, Princeton, NJ 08540-7219, USA}}
\newcommand{\Moore}{\affiliation{Gordon and Betty Moore Foundation, 1661 Page Mill Road, Palo Alto, CA 94304-1209, USA}}
\newcommand{\Durham}{\affiliation{Department of Physics, Durham University, South Road, Durham, DH1 3LE, UK}}
\newcommand{\JHU}{\affiliation{Department of Physics and Astronomy, The Johns Hopkins University, 3400 North Charles Street, Baltimore, MD 21218, USA}}
\newcommand{\Toronto}{\affiliation{David A.\ Dunlap Department of Astronomy and Astrophysics, University of Toronto,\\ 50 St.\ George Street, Toronto, Ontario, M5S 3H4 Canada}}
\newcommand{\Duke}{\affiliation{Department of Physics, Duke University, Campus Box 90305, Durham, NC 27708, USA}}

\title{SuperRAENN: A Semi-supervised Supernova Photometric Classification Pipeline Trained on Pan-STARRS1 Medium Deep Survey Supernovae}

\correspondingauthor{V. A. Villar}
\email{vav2110@columbia.edu}

\author[0000-0002-5814-4061]{V. Ashley Villar}
\CfA \Simons

\author[0000-0002-0832-2974]{Griffin~Hosseinzadeh}
\CfA

\author[0000-0002-9392-9681]{Edo~Berger}
\CfA

\author[0000-0002-0144-387X]{Michelle~Ntampaka}
\CfA\DSI

\author[0000-0002-6230-0151]{David~O.~Jones}
\UCSC

\author{Peter~Challis}
\CfA

\author[0000-0002-7706-5668]{Ryan~Chornock}
\CIERA

\author[0000-0001-7081-0082]{Maria~R.~Drout}
\Toronto\Carnegie

\author{Ryan~J.~Foley}
\UCSC

\author[0000-0002-1966-3942]{Robert~P.~Kirshner}
\Moore\CfA

\author[0000-0001-9454-4639]{Ragnhild~Lunnan}
\OKC

\author[0000-0003-4768-7586]{Raffaella~Margutti}
\CIERA

\author[0000-0002-0763-3885]{Dan~Milisavljevic}
\Purdue

\author{Nathan~Sanders}
\berkman

\author[0000-0001-8415-6720]{Yen-Chen~Pan}
\NAOJ

\author[0000-0002-4410-5387]{Armin~Rest}
\STScI\JHU

\author[0000-0002-4934-5849]{Daniel~M.~Scolnic}
\Duke

\author[0000-0002-7965-2815]{Eugene~Magnier}
\IfA

\author[0000-0001-9034-4402]{Nigel~Metcalfe}
\Durham

\author[0000-0002-1341-0952]{Richard~Wainscoat}
\IfA

\author[0000-0003-1989-4879]{Christopher~Waters}
\Princeton

\begin{abstract}
Automated classification of supernovae (SNe) based on optical photometric light curve information is essential in the upcoming era of wide-field time domain surveys, such as the Legacy Survey of Space and Time (LSST) conducted by the Rubin Observatory. Photometric classification can enable  real-time identification of interesting events for extended multi-wavelength follow-up, as well as archival population studies. Here we present the complete sample of \allsne\ ``SN-like" light curves (in g$_\mathrm{P1}$r$_\mathrm{P1}$i$_\mathrm{P1}$z$_\mathrm{P1}$) from the Pan-STARRS1 Medium-Deep Survey (PS1-MDS). The PS1-MDS is similar to the planned LSST Wide-Fast-Deep survey in terms of cadence, filters and depth, making this a useful training set for the community. Using this dataset, we train a novel semi-supervised machine learning algorithm to photometrically classify \newsne\ new SN-like light curves with host galaxy spectroscopic redshifts. Our algorithm consists of a random forest supervised classification step and a novel unsupervised step in which we introduce a recurrent autoencoder neural network (RAENN). Our final pipeline, dubbed {\tt SuperRAENN}, has an accuracy of 87\% across five SN classes (Type Ia, Ibc, II, IIn, SLSN-I). We find the highest accuracy rates for Type Ia SNe and SLSNe and the lowest for Type Ibc SNe. Our complete spectroscopically- and photometrically-classified samples breaks down into: \alliaper\% Type Ia (\allia\ objects), \alliipper\% Type II (\alliip\ objects), \alliinper\% Type IIn (\alliin\ objects), \allibcper\% Type Ibc (\allibc\ objects), and \allslsneper\% Type I SLSNe (\allslsne\ objects). Finally, we discuss how this algorithm can be modified for online LSST data streams.
\end{abstract}


\section{Introduction}
Time-domain astrophysics has entered a new era of large photometric datasets thanks to on-going and upcoming wide-field surveys, including Pan-STARRS (PS; \citealt{kaiser2010}), the Asteroid Terrestrial-impact Last Alert System (ATLAS; \citealt{jedicke2012atlas}), the All-Sky Automated Survey for SuperNovae (ASASSN; \citealt{shappee2014all}), the Zwicky Transient Facility (ZTF; \citealt{kulkarni2018zwicky}), the Legacy Survey of Space and Time (LSST; \citealt{ivezic2011large}), and the Roman Space Telescope \citep{spergel2015wide}. LSST, to be conducted by the Vera C. Rubin Observatory between 2023 and 2033, is expected to discover roughly one million SNe per year, a more than two orders of magnitude increase compared to the current rate. 

Historically, SNe and other optical transients have been classified primarily based on their optical spectra. Class labels are largely phenomenological, dependent on the presence of various elements in the photospheric-phase spectra (see e.g., \citealt{filippenko1997optical} for a review). SNe, for example, have historically been classified as Type I (equivalent to today's Type Ia) or Type II based on the absence or presence of strong hydrogen Balmer lines, respectively. As the number of events increased, further classes were created to account for the increased diversity (e.g., \citealt{uomoto1985peculiar}). Type Ib and Type Ic designations were created to indicate the presence and absence of helium, respectively. Today, semi-automated software such as {\tt GELATO} \citep{2008Harutyunyan}, {\tt SNID} \citep{blondin2007determining} and {\tt Superfit} \citep{howell2005gemini} are used to match SN spectra to a library of previously classified events to determine the spectroscopic class. More recently, \cite{2019Muthukrishna} utilized a convolutional neural network to classify SN spectra.

However, spectroscopic follow up remains an expensive endeavor, taking up to an hour on 8-meter class telescopes to classify a single object given the depth wide-field surveys can now achieve. As a result, only $\sim 10\%$ of the $\sim 10^4$ transients currently discovered each year are spectroscopically classified\footnote{Based on data from the public Open Supernova Catalog \citep{guillochon2017open} and the Transient Name Server.}. Spectroscopic follow up is not expected to significantly increase when the LSST commences, meaning that only $\sim 0.1\%$ of events will be spectroscopically classified.

Given the growing rate of discovery and limited spectroscopic resources, classification of transients based on their photometric light curves is becoming essential. Luckily, the phenomenological labels often correspond to unique underlying processes that are also encoded in the light curve behavior. For example, while Type Ia SNe are spectroscopically classified by strong Si II absorption and lack of hydrogen, these features distinctly originate from the thermonuclear detonations or deflagrations of carbon-oxygen white dwarfs, which also lead to specific light curve evolution \citep{hillebrandt2000type}. Generally, unique progenitor system and explosion mechanisms likely lead to other observable features, some of which are captured in broadband optical light curves. Said features allow transients to be classified into their traditional subclasses (based on spectroscopy and photometry) using \textit{only} their broadband, optical light curves.

There is a growing literature on light curve classifiers that rely on data-driven and machine learning algorithms. Most studies use \textit{supervised} learning, in which the training set consists of SNe with known classes (e.g., \citealt{lochner2016, 2017ApJ...837L..28C,boone2019,villar2019,moller2020supernnova}). However, SN classification can benefit from \textit{semi-supervised} methods, in which the training set contains both labelled and unlabelled SNe. The unlabelled set is used to better understand low-dimensional structure in the SN dataset to improve classification. \citet{richards2012semi}, for example, created a diffusion map (a nonlinear dimensionality reduction technique) based on light curve similarities in shape and color using unlabelled data from the Supernova Photometric Classification Challenge (SPCC; \citealt{kessler2010results}). They use the diffusion map to extract 120 nonlinear SN features from each labelled SN, which are then used to train a random forest classifier. More recently, \citet{pasquet2018deep} introduced the {\tt PELICAN} classifier, also trained on synthetic SPCC data. {\tt PELICAN} uses a convolutional autoencoder to encode nonlinear SN features and a set of fully connected neural network layers to classify the full set of simulated SPCC light curves as Ia or non-Ia SNe.

\defcitealias{hossen20}{H20}

Here we introduce a new semi-supervised classification method for SNe, which utilizes a recurrent autoencoder neural network (RAENN). This method is uniquely trained on real (rather than simulated) data from the Pan-STARRS1 Medium Deep Survey (PS1-MDS) and is optimized for general SN classification (as opposed to Ia versus non-Ia classification). Our method has been trained on a combination of \specsne\ spectroscopically-classified SNe and 2,328 additional SN-like events. We then use RAENN and hand-selected features with a random forest to classify the PS1-MDS sample of \newsne\ previously unclassified SN-like transients with host galaxy spectroscopic redshifts. We publish the full set of light curves and associated labels for community use. We present an open source code listed on the Python Package Index as {\tt SuperRAENN} \citep{villar_superraenn}. A companion paper, \citet[][hereafter \citetalias{hossen20}]{hossen20}, presents and compares photometric classifications of the same dataset using an independent classification method (following the supervised methods of our previous work in \citealt{villar2019}).

The paper is organized as follows. In \S\ref{sec:ps1}, we review the PS1-MDS and associated sample of SN-like transients. In \S\ref{sec:semisup} we introduce the RAENN architecture and training procedure. We present the classification results and discuss implications in \S\ref{sec:red} and \S\ref{sec:dis}, respectively. We conclude in~\S\ref{sec:con}. Throughout this paper, we assume a flat $\Lambda$CDM cosmology with $\Omega_M=0.307$, and $H_0 = 67.8$ km s$^{-1}$ Mpc$^{-1}$ \citep{ade2014planck}.

\begin{figure}[t]
\includegraphics[width=0.5\textwidth]{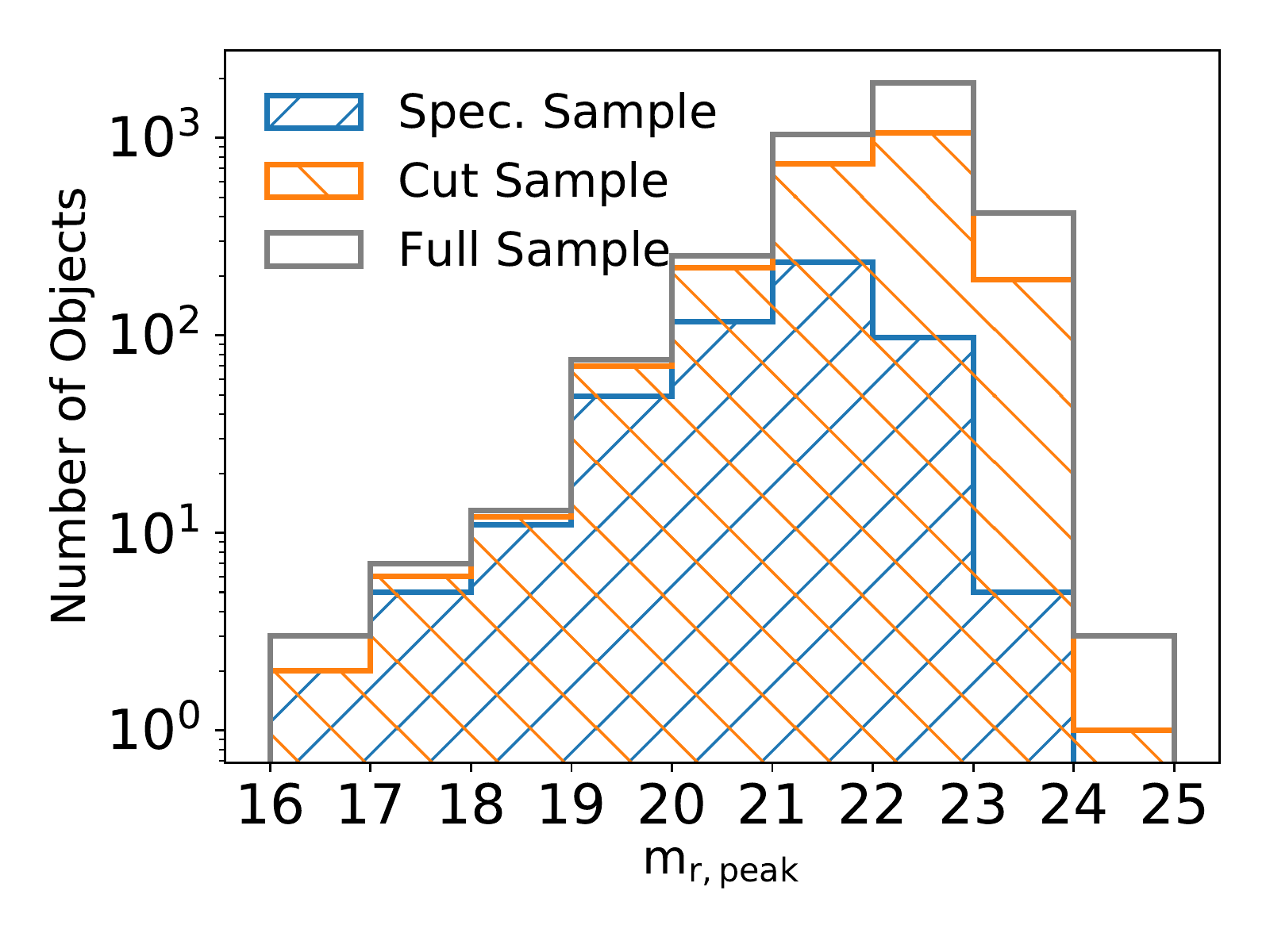}
\caption{Peak apparent $r$-band magnitude of the full SN-like dataset (grey), objects used in our unsupervised method (orange) and the spectroscopic sample (blue). The spectroscopic dataset is roughly one magnitude brighter than the full dataset. \label{fig:peaks}}
\end{figure}

\begin{figure}[t]
\includegraphics[width=0.5\textwidth]{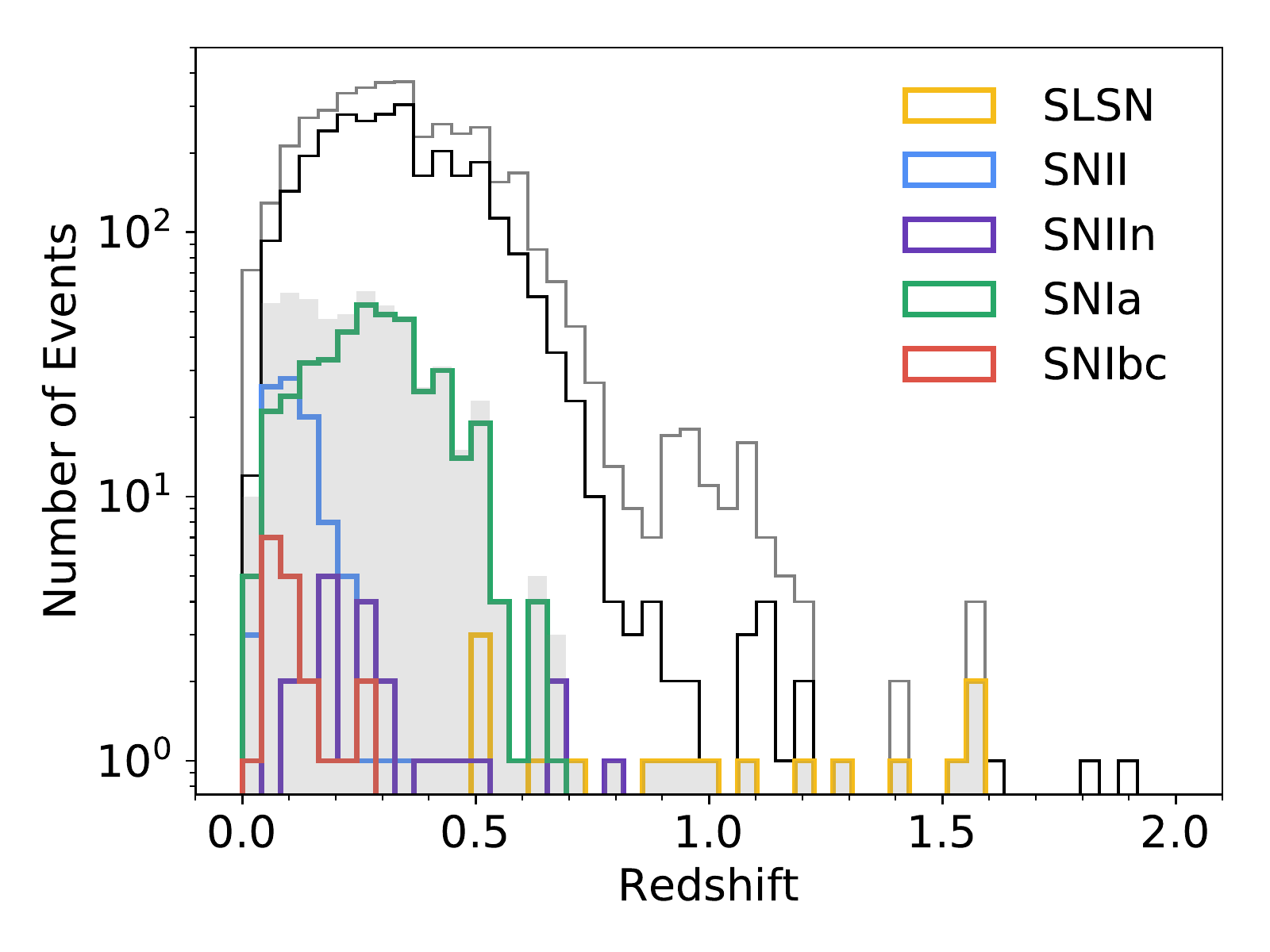}
\caption{Histogram of the redshifts of the full SN-like dataset (grey line; 4,055 objects), the subset of host redshift measurements for objects used in our unsupervised learning algorithm (black line;\unsupsne\ objects), and the subset with  spectroscopic classification (colored lines; \specsne\ objects). The shaded grey region represents the summed, spectroscopically classified objects. The full sample and spectroscopic distribution peak at $z\approx 0.25$, although the spectroscopic sample has an additional peak near $z \approx 0.1$. At $z\gtrsim 0.75$, the spectroscopic sample is limited to SLSNe. \label{fig:f1}}
\end{figure}

\section{The PS1-MDS Supernova Sample}\label{sec:ps1}

PS1 is a wide-field survey telescope located near the summit of Haleakala, Maui with a 1.8 m diameter primary mirror and a 1.4 gigapixel camera (GPC1) \citep{kaiser2010}. PS1-MDS, one of several PS1 surveys \citep{2016chambers}, was conducted between July 2009 and July 2014. It consisted of 10 single-pointing fields, each of approximately 7.1 deg$^2$, with a pixel-scale of $0.''25$.  The survey was conducted in five broadband filters \citep{stubbs2010precise,tonry2012pan} with a nominal cadence of $3$ days per filter in four filters ($g_\mathrm{P1}r_\mathrm{P1}i_\mathrm{P1}z_\mathrm{P1}$), and a $5\sigma$ limiting magnitude of $\approx 23.3$ per visit. In practice, \cite{scolnic2018complete} finds a cadence of roughly $6-7$ days per filter. In general, PS1-MDS observed a field in $g_\mathrm{P1}$ and $r_\mathrm{P1}$ on the same night, followed by $i_\mathrm{P1}$ and then $z_\mathrm{P1}$ on subsequent nights. PS1-MDS also included observations in the $y_\mathrm{P1}$-band, primarily near full moon and with a shallower $5\sigma$ limiting magnitude of $\approx 21.7$. Due to the poor cadence and shallow depth, we do not use the $y_{\mathrm{P1}}$ data here.

During its 5-year survey, PS1-MDS discovered \allsne\ SN-like objects, defined as events with at least three observersations with a signal-to-noise ratio (SNR) $>4$ in any filter and no previous detection within the survey \citep{jones2018measuring, jones2019foundation}. We obtain data for these events via the PS Data Processing System \citep{2016magnier,magnier2016panstarrs, waters2016panstarrs}. The photometric pipeline is based on {\tt photpipe} \citep{rest2005testing,rest2014cosmological} with improvements made in \cite{scolnic2018complete}. Images and templates, used for image subtraction, are re-sampled and aligned to match a ``skycell" in the PS1 sky tessellation. Image zeropoints are determined by comparing point spread function (PSF) photometry of stars to PS1 stellar catalogs \citep{2016chambers}. PS1 templates are convolved to match nightly images and then subtracted using {\tt HOTPANTS} \citep{becker2015hotpants}. For each event, a flux-weighted centroid is calculated and forced PSF photometry is performed at the centroid. Finally, a nightly zeropoint is applied. 

Of the \allsne\ SN-like objects, 4,090 host galaxies were targeted through a concerted observational effort. To identify the most likely host galaxy for each SN, we used the galaxy size and orientation-weighted R-parameter from \citet{sullivan2006rates}, as outlined in \citet{jones2017measuring}. The majority (3,321 objects) were observed using the Hectospec multifiber instrument on MMT \citep{fabricant2005hectospec,mink2007automating}. Additional host redshifts were obtained with the Anglo-Australian Telescope (AAT; 290 objects), the WIYN telescope (217 objects), and the Apache Point Observatory 3.5m telescope (APO; 5 objects). Host galaxies selected for follow-up were largely unbiased in terms of transient properties (e.g., we did not prioritize SNe based on luminosity, color or amount of additional followup). Additional host redshifts were obtained from archival survey data: 2dFGRS \citep{colless2003}, 6dFGS \citep{jones20096df}, DEEP2 \citep{newman2013deep2}, SDSS \citep{smee2013}, VIPERS \citep{scodeggio2018}, VIMOS \citep{le2005vimos}, WiggleZ \citep{blake2008wigglez} and zCOSMOS \citep{lilly2009zcosmos}.

We use the {\tt RVSAO} package \citep{kurtz1998rvsao} to determine the spectroscopic redshifts through cross-correlation with galaxy templates. We use the standard {\tt RVSAO} galaxy templates (including spiral and elliptical galaxies and quasars), as well as galaxy templates provided by SDSS \citep{sdss5}\footnote{\url{http://classic.sdss.org/dr5/algorithms/spectemplates/}}. We quantify the quality of the template matches using the \citet{tonry1979survey} cross-correlation parameter, $R_\mathrm{CC}$. Following \citet{jones2017measuring}, we remove host galaxies with $R_\mathrm{CC}<4$, ensuring that the vast majority ($\approx98$\%) of the remaining galaxies have accurate redshift measurements. This cut removes 1,084 SNe with redshift measurements.

To ensure that our final set of redshift measurements is robust, we identify a subset of spectra to be manually validated. Of the remaining redshifts which we initially estimate using {\tt RVSAO}, we accept the redshift of the best-matching template without visual inspection if the median redshift estimate across templates is equal to both the most-likely redshift and the mode of the template matches \textit{and} more than two templates match this redshift estimate. We (VAV and GH) visually inspected $\sim600$ redshift spectra to ensure that our final redshift estimates are as accurate as possible. In total, 2,487 redshifts (of 3,056 redshift estimates with $R_\mathrm{CC}\ge4$) match the most-likely redshift provided by {\tt RVSAO}. Of the remaining hosts, we remove 393 redshift estimates which we could not validated manually. A total of 145 redshifts ($\sim4$\%) which were measured manually do not match the median or mode of the {\tt RVSAO} redshift estimates. The galaxy spectra and further details are presented in \citetalias{hossen20}.

We additionally remove events with $z<0.005$, which are unlikely to be SNe given the peak absolute magnitudes (e.g., \citealt{2010ATel.2680....1C}). We visually inspect the light curves which have quasar-like hosts (based on template matching) or which overlap with the host galaxy's center. We remove events which are clearly variable over multiple seasons and lack a transient spectrum. Our final sample includes \unsupsne\ transients with redshifts measurements (from the hosts or transients themselves), including spectroscopically-identified SNe. 

\subsection{Spectroscopic versus Photometric SN Sample}

Approximately 10\% of the PS1-MDS transients were spectroscopically observed in real time throughout the survey, without a specific selection function (although brighter objects were more likely to be targeted). For this work, we limit our spectroscopic sample (\specsne\ objects) to five potential classes: 

\begin{enumerate}
\item Type I SLSNe (\numslsne\ objects), which are thought to arise from the birth of highly magnetized neutron stars \citep{quimby2007sn,chomiuk2011pan,nicholl2017magnetar}
\item Type II SNe (\numiip\ objects; including Type IIP and Type IIL SNe\footnote{Type IIP and Type IL are thought to arise from the same progenitor population. See e.g., \cite{sanders2015toward}}),  which arise from red supergiant progenitors
\item Type IIn SNe (\numiin\ objects), powered by the interaction the SN ejecta with pre-existing circumstellar material (e.g. \citealt{smith2014sn})
\item Type Ia SNe (\numia\ objects), which are the thermonuclear explosions of white dwarfs
\item Type Ibc SNe (\numibc\ objects), which arise from the core-collapse of massive stars that have lost their hydrogen (Ib) and helium (Ic) envelopes. Due to the small sample size we consider Type Ib and Type Ic SNe as a single class. 
\end{enumerate}

The SLSN and Type Ia SN light curve samples have been previously published in \citet{lunnan2018hydrogen} and \citet{jones2017measuring}, respectively. Model fits to the Type II light curves were presented in \citet{sanders2015toward}. For four objects, the transient spectra yield a reliable redshift but an ambiguous classification. A fifth object, PSc130816, has previously been identified as both a Type IIP/L SN \citep{sanders2015toward} and a Type IIn SN \citep{drout2016peculiar}. We do not include these five objects in our spectroscopic sample. An additional 15 objects are spectroscopically identified but do not fall in on of our five classes, including two tidal disruption events (TDEs), a lensed Type Ia, a Type Ibn, a Type Iax and ten fast evolving luminous transients (FELTs). All except the TDEs are included in our photometric sample for training purposes, but not included in our spectroscopic sample. These objects are discussed in more detailed in \S\ref{sec:dis1}

Our photometric sample contains \newsne\ objects with host galaxy spectroscopic redshifts, that are independent of the \specsne\ SNe which are spectroscopically classified. We refer to the union of the photometric and spectroscopic samples (the full set of \unsupsne\ events), as the ``complete" photometric dataset.  We summarize the PS1-MDS SN-like objects, their associated hosts and redshift information in Table~\ref{tab:one}. We also specify which SNe are used in the supervised/unsupervised portions of our classification algorithm.

Our spectroscopic dataset is brighter than our complete photometric dataset. As shown in Figure \ref{fig:peaks}, the spectroscopic sample has a median peak $r$-band magnitude of $\sim-21$ mag, about 1 magnitude brighter than the photometric sample. We directly compare the redshift distributions in Figure \ref{fig:f1}. The spectroscopic sample peaks at a slightly lower redshift compared to the photometric dataset($z\approx0.27$ versus $z\approx0.35$), with a tail extending to $z\approx1.0$. The lack of confident high-redshift measurements is likely due to the key spectroscopic lines shifting out of the optical and due to the peak absolute magnitudes of most SNe falling below our limiting magnitude. The mismatch between the spectroscopic and photometric samples may translate to biases in our classification pipeline, which we explore in more detail in \S~\ref{sec:dis}. The complete $griz$ light curves of our sample are available on Zenodo \citep{villar_lc_2020}. 

\begin{figure}
\includegraphics[width=0.5\textwidth]{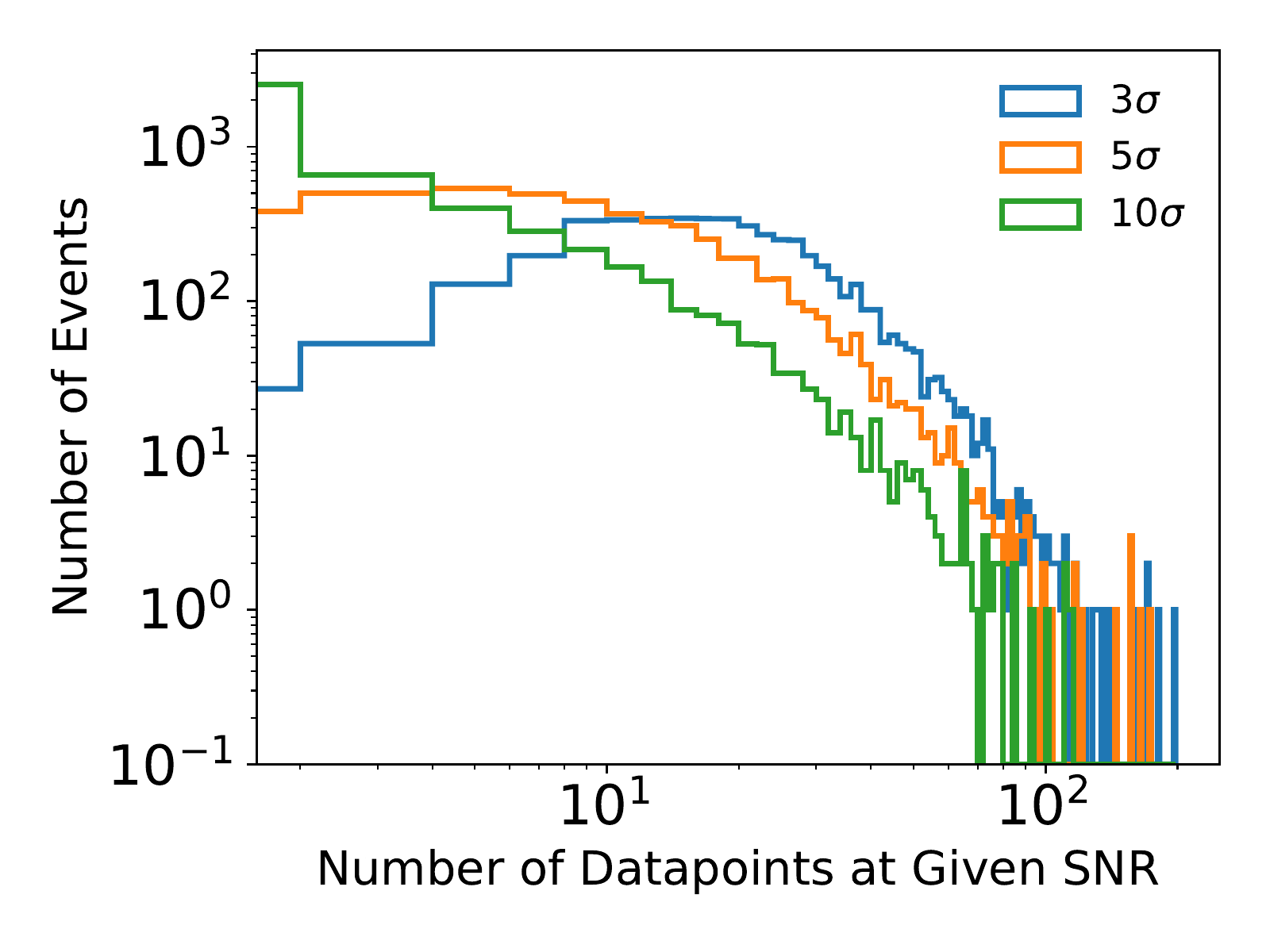}
\caption{ Histogram of the number of SN light curves with $N$ data points with SNR of $\ge 3$ (blue), $\ge 5$ (orange), and $\ge 10$ (green) from the complete sample of SN-like objects (\allsne\ events).  Most events have $\approx10-20$ $3\sigma$ data points, with only a handful having $>100$ points. \label{fig:sig}}
\end{figure}

We explore the overall data quality of our sample in Figure~\ref{fig:sig}, finding that the majority of events have $\sim20$ data points across all filters with signal-to-noise ratios   of $\gtrsim 3$. Given a typical SN duration of a month and our typical cadence of a few days, we expect the majority (but not all) SNe to have fairly complete light curves.

\section{A Semi-supervised Classification Pipeline}\label{sec:semisup}

About 10\% of our SN sample is spectroscopically classified. Traditional supervised classification methods are strictly limited to this subset of our data, as they require labelled SN examples. However, information about SN subtypes exists as substructure in the unlabelled dataset as well. For example, SN classes may be clustered in duration and luminosity (e.g., \citealt{kasliwal2012systematically,villar2017theoretical}). Because we would like to leverage the information in both the labelled and unlabelled subsets of the training set, we use a recurrent autoencoder neural network (RAENN) paired with a random forest classifier for a semi-supervised classification approach. In this section, we describe the complete algorithm and training process. 

Our pipeline is composed of three steps: (1) a pre-processing and interpolation step using Gaussian processes (GP); (2) an unsupervised step in which we train a RAENN on the complete photometric set (labelled and unlabelled); and (3) a supervised step in which we train a random forest on the spectroscopically labelled set of SNe. The complete pipeline, dubbed {\tt SuperRAENN} \citep{villar_superraenn}, is available via GitHub\footnote{\url{https://github.com/villrv/SuperRAENN}}.

\subsection{Pre-processing with Gaussian Processes}

\begin{figure*}
\includegraphics[width=\textwidth]{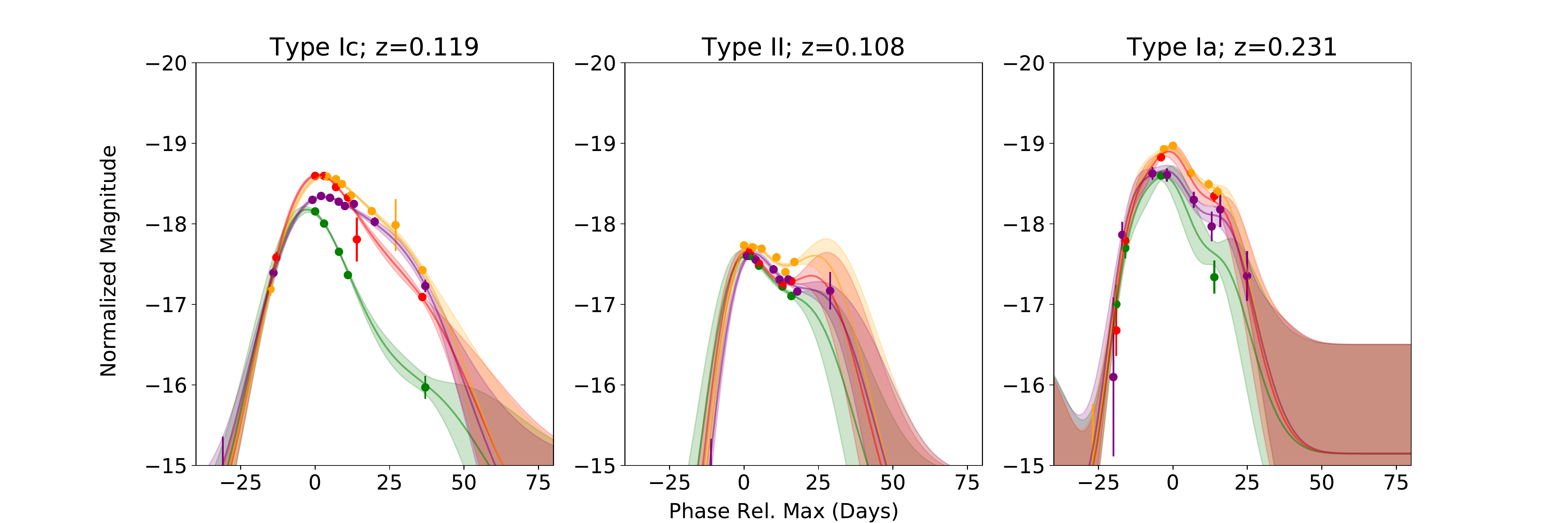}
\caption{Examples of three spectroscopically classified SNe and their associated GP-interpolated light curves in the four PS1 filters ($g$: green; $r$: red; $i$: orange; $z$: purple). Solid lines represent the mean GP prediction, while the shaded regions represent the $1\sigma$ estimated uncertainties.  \label{fig:gp}}
\end{figure*}

We generate and pre-process absolute magnitude light curves before extracting features. We correct each light curve for Milky Way reddening using the extinction map of  \citet{schlafly2011measuring}.  We estimate and normalize the absolute magnitude using the measured host redshift:
\begin{equation}
\begin{split}
    M_\mathrm{norm} = m - 5\log_{10}(d_\mathrm{L}/10\mathrm{pc})\\ + 2.5\log_{10}(1+z) - m_\mathrm{lim} - A_\lambda
\end{split}
\end{equation}
where $m_\mathrm{lim}$ is a chosen limiting magnitude, which we take to be $m_\mathrm{lim}=25$. This value is dimmer than the $5\sigma$-limiting magnitude of PS1-MDS. We choose a dimmer magnitude to ensure that even marginal detections will be included in the light curve. We perform the re-normalization so that the GP mean will be zero (i.e., the light curve will be zero when no data is available). Finally, we correct all light curves for time-dilation based on the measured redshifts. We do not attempt to make a wavelength-dependent $k$-correction to the rest-frame data given the complicated, diverse, and time-evolving spectral energy distributions (SEDs) of the various SN types.

We do not correct the SN light curves for host galaxy reddening. The intrinsic reddening of SNe adds an additional scatter in our feature space. Correcting for host galaxy reddening would require estimating both the color excess and dust law, which is not possible given our current dataset. 

The PS1-MDS light curves are irregularly sampled across the four filters (see \S\ref{sec:ps1} for the PS1 observing strategy). The architecture of the RAENN does \textit{not} require uniformly sampled light curves. However, it does require that each observation is made in all four filters. For example, if an observation is made in $g$-band, we need to provide interpolated values for $riz$-bands for that time. 

To interpolate the $griz$ light curves, we fit a GP using the open-source Python package {\tt George} \citep{foreman2015george}. GPs are a non-parametric model that has been previously used to interpolate and classify SN light curves (see e.g., \citealt{lochner2016,revsbech2018,boone2019}). GPs define a prior over a family of functions, which is then conditioned on the light curve observations.  A key assumption is that the posterior distribution describing the light curve is Gaussian, described by a mean, $\mu(\vec{t})$, and a covariance matrix, $\Sigma(\vec{t})$, given by $\Sigma_{i,j}=\kappa(\vec{x_i},\vec{x_j})$ with kernel $\kappa$. We use a 2D squared exponential kernel to simultaneously fit all four filtered light curves:

\begin{equation}
\begin{split}
\kappa(\vec{t_i}\vec{t_j}\vec{f_i}\vec{f_j}; \sigma, l_{t}l_{f})= \sigma^2 \exp\Big[-\frac{(t_i-t_j)^2}{2l_t^2}\Big]\\
\times\exp\Big[-\frac{(f_i-f_j)^2}{2l_f^2}\Big]
\end{split}
\end{equation}
where $f$ is an integer between 0 and 3 that represents the $griz$ filters, and the parameters $l_t$ and $l_f$ are characteristic correlation length scales in time and filter integer, respectively. This fitting process accounts for the measured data uncertainties, making it robust to low-confidence outliers.

We independently optimize the kernel parameters for each SN using the {\tt minimize} function implemented in {\tt scipy}, with initial values of $l_t=100$ days and $l_f=1$. We find that our choice of initialization values has little effect on the resulting best fit.  We find that $l_t$ is typically about one week, and $l_f$ is typically $2-3$, indicating that the filters are highly correlated. Examples of the GP interpolation for Type Ia, Type Ic and Type II SNe are shown in Figure~\ref{fig:gp}. The GP is able to produce reasonable interpolated light curves even in cases with sparse and noisy data and provide reasonable error estimates. 

A similar GP method was implemented by \citet{boone2019} to classify a variety of SN types in the Photometric LSST Astronomical Time-series Classification (PLAsTiCC; \citealt{allam2018photometric,kessler2019models}) dataset. Instead of an integer, \citealt{boone2019} used the rest frame central wavelength of each filter for each object. We avoid this added layer of complexity because the $k$-corrections and time-evolving SN spectral energy distribution (SED) change the weighted central filter wavelength. However, the simple 2D kernel still allows the four bands to share mutual information.


Our light curves contain several years of data, most of which are non-detections. To limit our input data, we keep datapoint (of any significance) within 100 days of peak flux (in whichever filter is brightest). For ease of optimization, the light curves need to contain the same number of data points. The data must be a consistent size during the back-propagation step of optimization for the RAENN for each iteration (see next section). Our longest light curve contains 169 data points, so we pad all light curves to match this length. We do so by appending a value dimmer than the estimated absolute limiting flux (we use $m_\mathrm{lim}=25$) to 100 days after the last detection in the light curve. 

We note that using luminosity-based light curves (rather than magnitudes) is an alternative pre-processing choice. Luminosity-based light curves would remove the need to re-normalize the light curves to a chosen limiting magnitude.  We find that using luminosity-based light curves results in worse performance of the RAENN, likely due to the orders-of-magnitude differences in scale between events.

\subsection{Unsupervised Learning: A Recurrent Autoencoder Neural Network (RAENN)}

\begin{figure*}
\includegraphics[width=\textwidth]{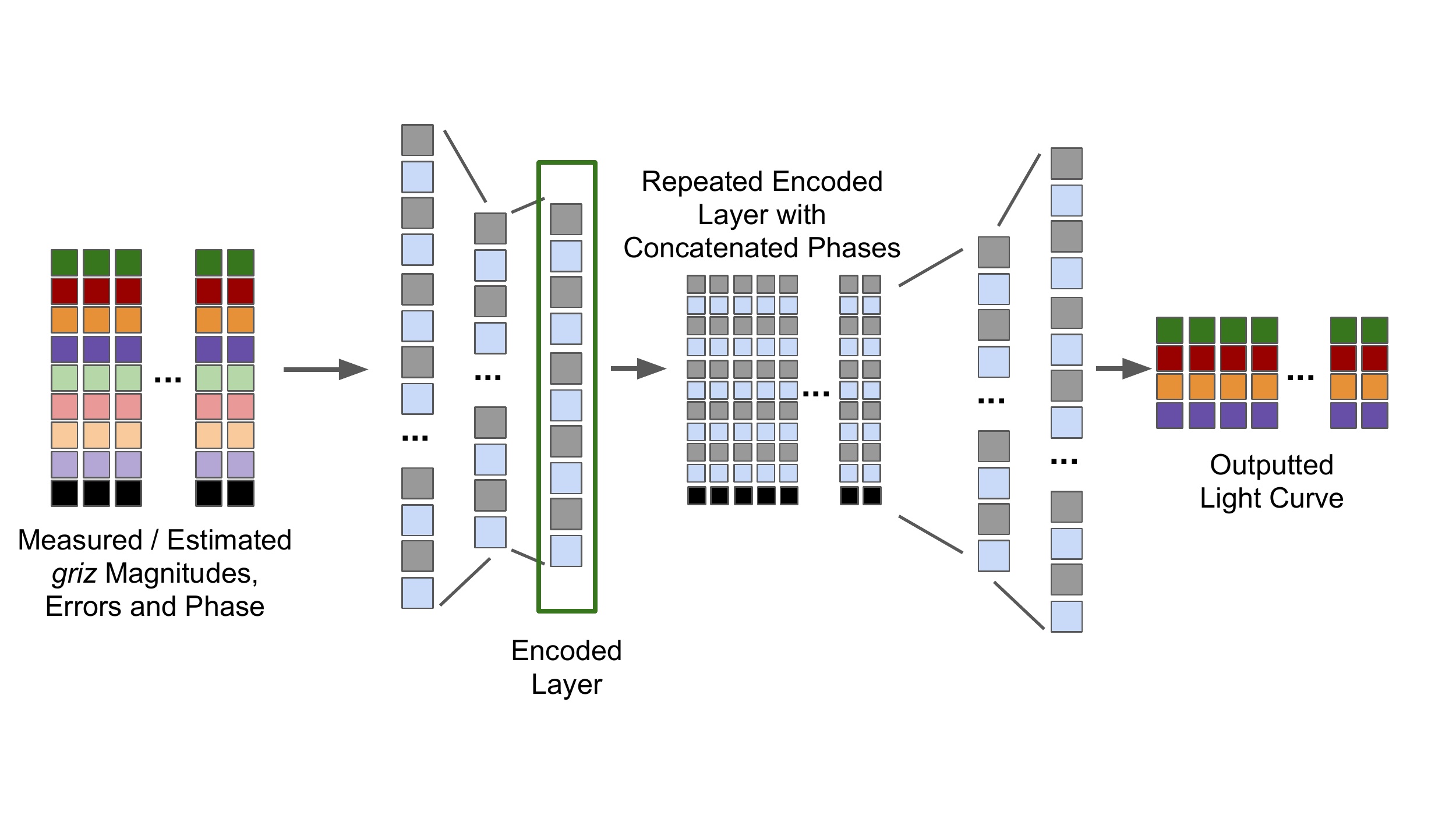}
\caption{Diagram of the RAENN architecture. The pre-processed GP-interpolated light curves are fed into the encoder, which encodes the light curve into an encoding vector. This vector is then repeated, and new time values are appended to each copy. The final light curve is then predicted at each new time value. The RAENN is trained by comparing the input light curve with the predicted light curve. The values from the encoded layer are inputted into the random forest as features and used to classify the SN light curves. \label{fig:nnd}}
\end{figure*}

To extract unique features from the complete (unlabelled and labelled) PS1-MDS photometric sample, we construct a RAENN, inspired by the work of \citet{naul2018recurrent}, who uses a similar method to classify variable stars.

Neural networks are a class of machine learning algorithms that use many latent layers to model complex functions.  These and other machine learning algorithms are becoming increasingly common in astronomy (see \citealt{ntampaka2019role} for an overview). Autoencoders (AEs, \citealt{kramer1991nonlinear}) are a class of neural network architectures that learn a compressed representation of input data. By training an AE to return the original data given a limited set of variables, it learns an ``encoded" version of the data. 

In astrophysics, AEs have been used for feature-learning in galaxy spectral energy distributions (SEDs, \citealt{frontera2017unsupervised}), image de-noising \citep{ma2018radio,lucas2018recovering}, and event classification \citep{naul2018recurrent,pasquet2019pelican}. AEs are also increasingly being used in the astrophysics literature for dimensionality reduction (see e.g., \citealt{ralph2019radio} and \citealt{portillo2020dimensionality} for recent examples). 

Here, our model is designed to address several concerns of SN light curves: (1) the temporal irregularity of data; (2) data across multiple filters; and (3) streaming data that update on a given cadence. The last point is not a concern for our PS1-MDS archival dataset, but it will become important as LSST comes online and discovers thousands of SNe nightly.

The RAENN uses the GP light curves as input, by codifying the light curves as matrices of size $9\times T_0$, where $T_0=169$, as described in the previous section. The $9$ values are: one time value, relative to maximum (in whichever filter is brightest); four magitude values ($griz$) at that time; and four magnitude uncertainties. Recall that the magnitude values are either measured or estimated from the GP. For the uncertainties, we use the $1\sigma$ errors for the measured points. For the GP points, we use a large error of 1 mag. We note that the GP produces estimates errors, but we find that, in practice, using this larger error bar leads to better performance. We leave exploration of utilizing error bars to future work. We emphasize that, while $T_0=169$ for training, the RAENN architecture allows a user to input a light curve of any size without needing to pad the light curve. 

The RAENN architecture is divided into an encoder and a decoder. Our encoder is a series of fully-connected layers that decrease in size until the final encoded layer with size $N_E$ (i.e., the number of neurons used to fully encode the SN light curve). We note that $N_E$ is a free parameter of our model that needs to be optimized. Similarly, the fully-connected layer has $N_N$ neurons, where $N_N>N_E$ and is also a tunable parameter. Following the encoded layer, the decoder half of the architecture mirrors the encoder with increasing layer sizes.

A novel feature of our architecture is the inclusion of a repeat layer immediately after our final encoding layer (the layer of size $N_E$). In this layer, we repeat the encoded version of the light curve $T_N$ times. To each copy, we append the time of each data point, relative to peak brightness in one filter (whichever filter happens to be brightest). One way to view the purpose of this layer is to imagine the autoencoder as two functions. The first function (the encoder) takes in the original data points, including observation times, and outputs a set of $N_E$ values. This is similar to the idea of taking a light curve and fitting it to a model with $
N_E$ free parameters. A second function (the decoder) takes in a set of $N_E$ values and $T_N$ times to generate a light curve at the $T_N$ times. This architecture allows us to generate a light curve at different $T_N$ times; e.g., interpolated or extrapolated light curves, which is further explored in \S\ref{sec:dis}. In this work, we choose $T_N=T_0$; namely, we repeat the encoded values to match the original light curve length. 

Our autoencoder utilizes gated recurrent neurons (GRUs; \citealt{cho2014learning,rumelhart1988learning}). In addition to the typical hidden weights that are optimized during training, recurrent neurons have additional weights that act as ``memory'' of previous input. GRUs in particular utilize an \textit{update} value (called a gate) and a \textit{reset} gate. The values of these neurons determine how the current and previous input affect the value of the output. With each light curve data point, the gates become updated with new information that informs the next prediction. This class of neurons is useful for our light curves with various numbers of observed data points. Our GRU neurons use the $\tanh$ activation functions with a hard sigmoid for the gate activation function. 

Our RAENN is implemented in {\tt Keras} \citep{chollet2015} with a Tensorflow backend \citep{abadi2016tensorflow}. A diagram of the architecture is shown in Figure~\ref{fig:nnd}, and is outlined as follows:

\begin{enumerate}
    \item \textbf{Input Layer}: Input light curve of size $T_0\times 9$ with each $griz$ data point labelled with a time (1 value) in days relative to light curve peak (4 values) and an uncertainty (4 values). 
    \item \textbf{Encoding Layer}: Encoding layer with $N_N$ neurons, where $N_N$ is a hyperparameter. 
    \item \textbf{Encoded Layer}: Encoded light curve with $N_E$ neurons, where $N_E$ is a hyperparameter.
    \item \textbf{Repeat Layer}: Layer to repeat encoded light curve to match with new time-array, with size $T_0\times N_E$.
    \item \textbf{Concatenate Layer}: Layer to concatenate new times to encoded light curve, with size $T_0\times (N_E+1)$. 
    \item \textbf{Decoding Layer}: Decoding layer with $N_N$ neurons.
    \item \textbf{Decoded Layer}: $T_0\times 4$ decoded $griz$ light curve.
\end{enumerate}

To optimize the free parameters (the weights) of the RAENN model, we must define a \textit{loss function}. Our loss function is a simple mean square error function:
\begin{equation}
    \mathcal{L}=
\sum_{i=0}^N\frac{\Big[F_{i,\mathrm{True}}(t,f) - F_{i,\mathrm{Predicted}}(t,f)\Big]^2}{N},
\end{equation}
where $F$ is the SN flux as a function of time $t$ and filter $f$. Although we feed unvertainties into the network, we find that excluding flux errors in our loss function substantially improved the ability of the RAENN to match the input light curves. We minimize our loss function using the gradient descent-based optimizer, {\tt Adam} \citep{kingma2014adam}, finding an optimal learning rate of $10^{-4}$, which is a typical value.

\begin{figure*}
\includegraphics[width=0.98\textwidth]{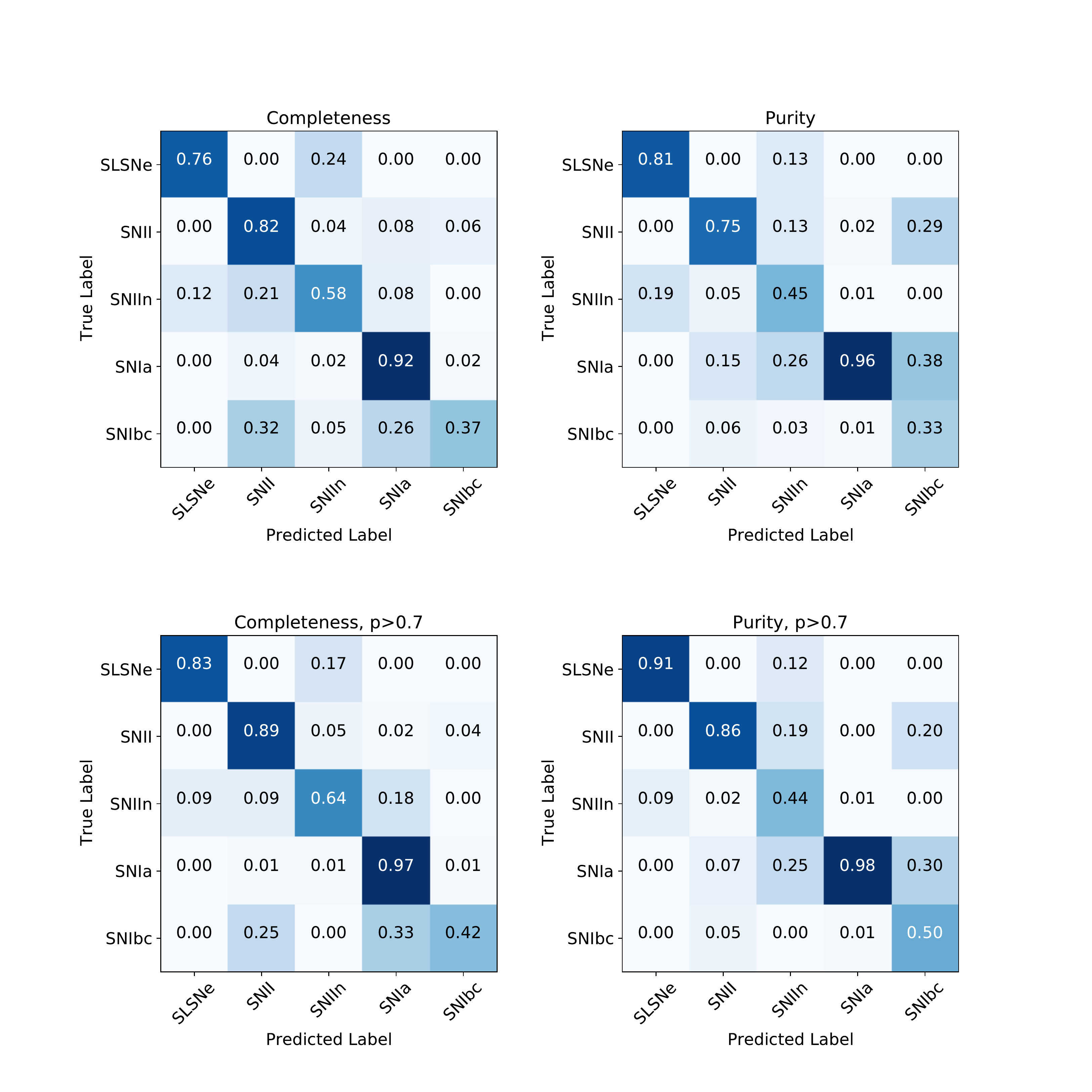}
\caption{Confusion matrices for the full set of \specsne\ spectroscopically classified SNe. In the bottom panel, we include only objects where the maximum probability is $\ge0.7$ (438 events). \textit{Left panels:} Completeness-based confusion matrices, in which each row is normalized to equal one. Completeness quantifies how much of a spectroscopic class the classifier has correctly classified. \textit{Right panels:} Purity-based confusion matrices, in which each column is normalize to equal one. Purity quantifies how much a photometric class is comprised of the true spectroscopic class. By restricting our classes to the high-confidence objects (bottom panels), both our completeness and purity increase. \label{fig:conf}}
\end{figure*} 

\begin{figure*}
\centering
\includegraphics[width=0.7\textwidth]{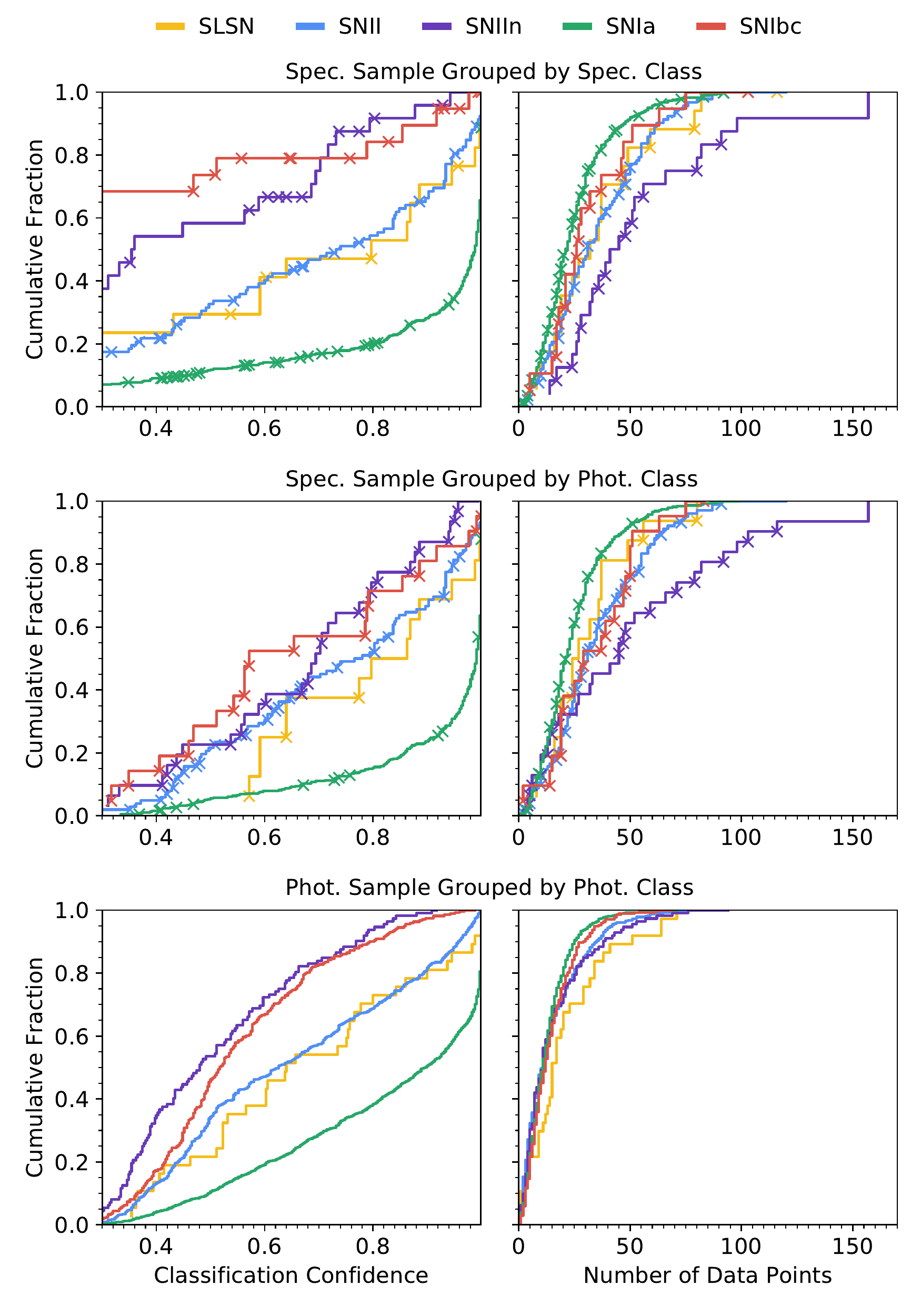}
\caption{\textit{Top:} Cumulative fraction of the spectroscopic SN sample as a function of classification confidence (left) and number of $>5\sigma$ data points (right), grouped by spectroscopic class. Misclassifications are marked with an ``x". \textit{Middle:} Cumulative fraction of the spectroscopic SN sample, grouped by photometrically-identified class. As expected, most misclassifications occur at low-confidence. At our chosen high-confidence cutoff ($p>0.7$), we find that the samples are largely pure. \textit{Bottom:} Cumulative fraction of the photometric SN sample, grouped by photometrically-identified class. The distributions based on classification confidence follow a similar trend to those seen in the spectroscopic sample, with Type Ia SNe and SLSNe having the highest fraction of high-confidence events. However, the photometric set has significantly more points on average when compared to the spectroscopic dataset.  \label{fig:cumula}}
\end{figure*}

We randomly split our unlabeled dataset into training (2/3) and test (1/3) sets. We optimize the number of neurons in both the encoding and decoding layers (fixed to be the same number, $N_N$) and the number of encoding neurons ($N_E$) through a grid search, allowing $N_N$ to vary from 20 to 160 in intervals of 20, and $N_E$ to vary between 2 and 24 in intervals of 2. We find that, when optimizing over final classification F1-score (defined below), purity and completeness, our results are relatively insensitive to $N_E$ and $N_N$ for values of $N_E\sim10$ and $N_N\sim100$. For our final model, we use $N_E=8$ and $N_N=120$, which is \textit{not} our optimal model but a \textit{representative} model. Utilizing our optimal model without creating a valid test set (in addition to a training and validation set) would likely overestimate performance. Given our limited dataset, we are unable to properly optimize our hyperparameters and thus present representative results. We note that $N_N$ is slightly below the maximum number of data points in our set of light curves (where the longest light curve has 169 observed data points). The number of encoding neurons $N_E$ is similar to the number of free parameters for the analytical model used in \citet{villar2019} to capture the shape of a single-filter SN light curve.

We contrast our architecture with methods from \citet{naul2018recurrent} and \citet{pasquet2019pelican}, who present similar methodologies. \citet{naul2018recurrent} uses a similar GRU-based RAENN to classify variable stars with unevenly sampled light curves in one filter from the All Sky Automated Survey Catalog of Variable stars \citep{pojmanski2002all}. The flux and time since last observation ($\Delta t$) is sequentially read into the recurrent layers. The same time array is fed into the decoder for output. In our case, we feed in a time series across four filters and give a time array relative to peak rather than relative to the previous data point. This is more natural in our problem, in which the SNe have a clear beginning and end, versus the periodic signals of variable stars.  Additionally, our architecture allows us to give the decoder a different time series to allow for interpolation or extrapolation of the data. 

\citet{pasquet2019pelican} uses a semi-supervised method to classify simulated SN light curves from the SPCC \citep{kessler2010results}. They use an AE with convolutional layers by transforming the light curves into ``light curve images'' (see \citealt{pasquet2018deep}). Rather than interpolate the light curves, \citet{pasquet2019pelican} applies a mask to filters that are missing data at a certain time. In contrast, we interpolate our light curves but assign interpolated values a large uncertainty of 1 mag, as explained above. We found that the method of transforming light curves into images and masking across four filters led to unstable training and poorer performance. This is likely due to the large data gaps in the {\it real} PS1-MDS light curves, compared to the high-cadence (2-days for each filter) {\it simulated} light curves of SPCC.  Since the LSST data are expected to more closely resemble the PS1-MDS light curves than the SPCC simulated events, we expect our method to be more robust in a real-life application.

\begin{figure}
\includegraphics[width=0.45\textwidth]{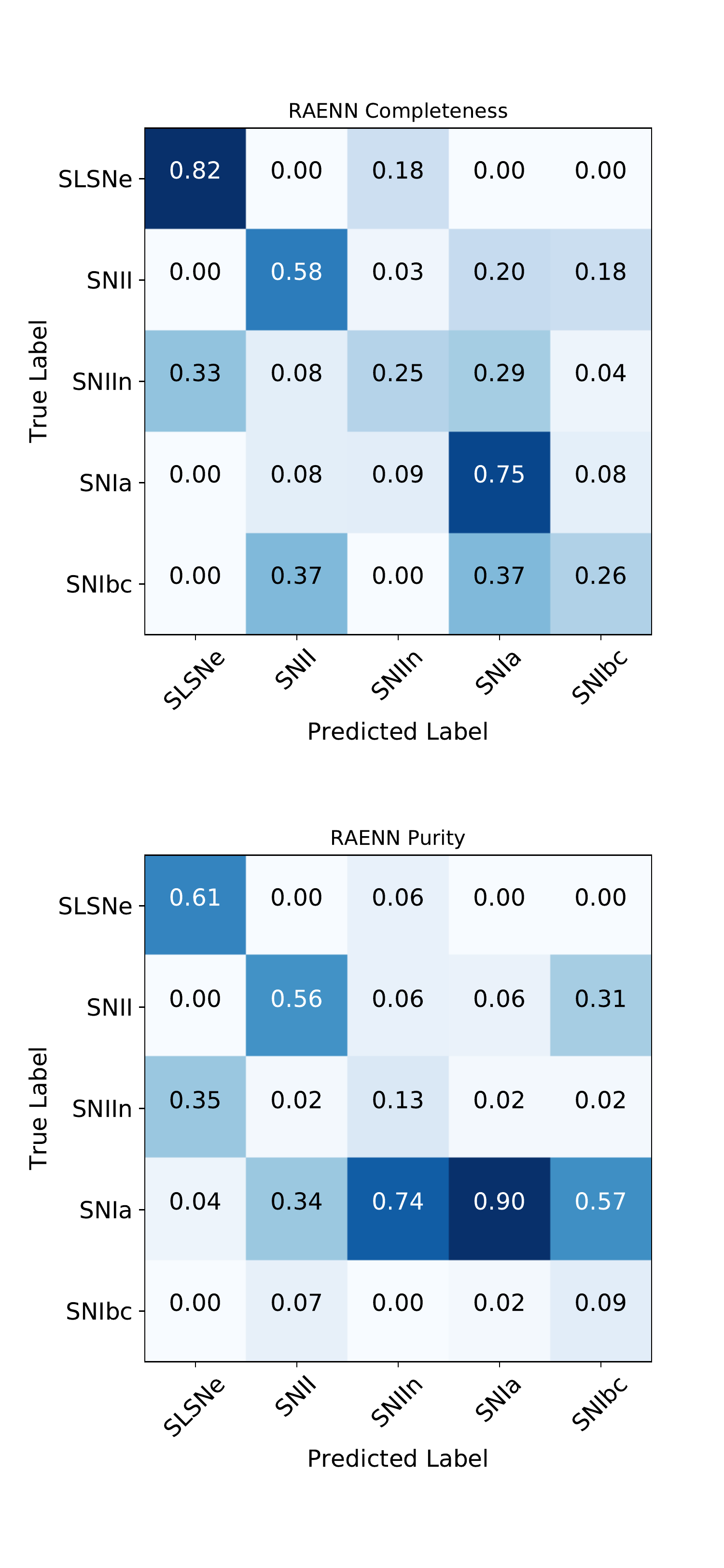}
\caption{ Completeness and purity confusion matrices, generated from classifying the spectroscopic dataset using only RAENN features and leave-one-out cross-validation. Even without additional features, the classifier performs similarly to other simulation-based classifiers such as those presented in \citet{muthukrishna2019rapid} and \citet{boone2019}.  \label{fig:conf21}}
\end{figure}

\subsection{Supervised Learning: Random Forest Classifier}

As a final step, we use the encoded light curves as features for a supervised classification method. Following \citet{villar2019}, we train a random forest (RF) classifier on the PS1-MDS spectroscopically classified SNe, including the RAENN encodings as features. 

In addition to the encoding (8 features), we use the following 36 features based on the GP-interpolated light curves: 

\begin{itemize}
    \item The $griz$ rise times in the rest frame, calculated 1, 2, and 3 mag below peak (12 features). 
    \item The $griz$ decline times in the rest frame, calculated 1-, 2- and 3-magnitudes below peak (12 features).
    \item The $griz$ peak absolute magnitudes (4 features) 
    \item The median $griz$ slope between 10 and 30 days post-peak in observer frame. This area was chosen by eye to specifically help the model differentiate between Type II and Type Ibc SNe (4 features).
    \item The integral of $griz$ light curves (4 features).
\end{itemize}

We measure these values from the GP-interpolated light curves rather than the decoded light curves. The decoded light curves are, at best, approximations of the GP-interpolated light curves. Therefore, using them would only result in noisier features. The decoded light curves are necessary, however, as a means to train the RAENN to extract the $N_E$ encoding neuron values. We note that for some features, e.g., the rise and decline times, the feature values are heavily dependent on the GP extrapolation in cases where there is no measured data. Including GP errors in the supervised portion of our analysis could help capture this intrinsic uncertainty in the underlying light curve, but we leave that exploration to future work.

These features were chosen through trial-and-error while optimizing classification accuracy. We find that inclusion of all features leads to our optimal classification accuracy, although we do explore how well our classifier performs with the RAENN features alone in the following section.

We pass these features through a RF classifier, utilizing 350 trees in the random forest and the Gini-information criterion. The number of trees was determined based on trial-and-error optimization. To counteract the imbalance across the five spectroscopic classes, we tested several algorithms to generate synthetic data to augment our training set. Following \citealt{villar2019}, we use a Synthetic Minority Over-sampling Technique (SMOTE; \citealt{chawla2002smote}) and a multivariate Gaussian (MVG) fit. We additionally test using a Kernel Density Estimate (KDE) of the training set, using a Gaussian kernel with bandwidth equal to 0.2 (or 20\% of the whitened feature standard deviation). We find that the MVG with a halved covariance matrix performs best. We test our classifier using leave-one-out cross validation, in which we remove one SN from the sample, oversample the remaining objects by generating new objects using the MVG, and then apply the trained RF to the single, removed event and recording the result. For each object, our RF reports probabilities associated to each class, which are calculated using the fraction of trees which vote for each class. We take the class with the highest probability as the predicted SN type.

\section{Classification Results}\label{sec:red}

\begin{figure}
\includegraphics[width=0.45\textwidth]{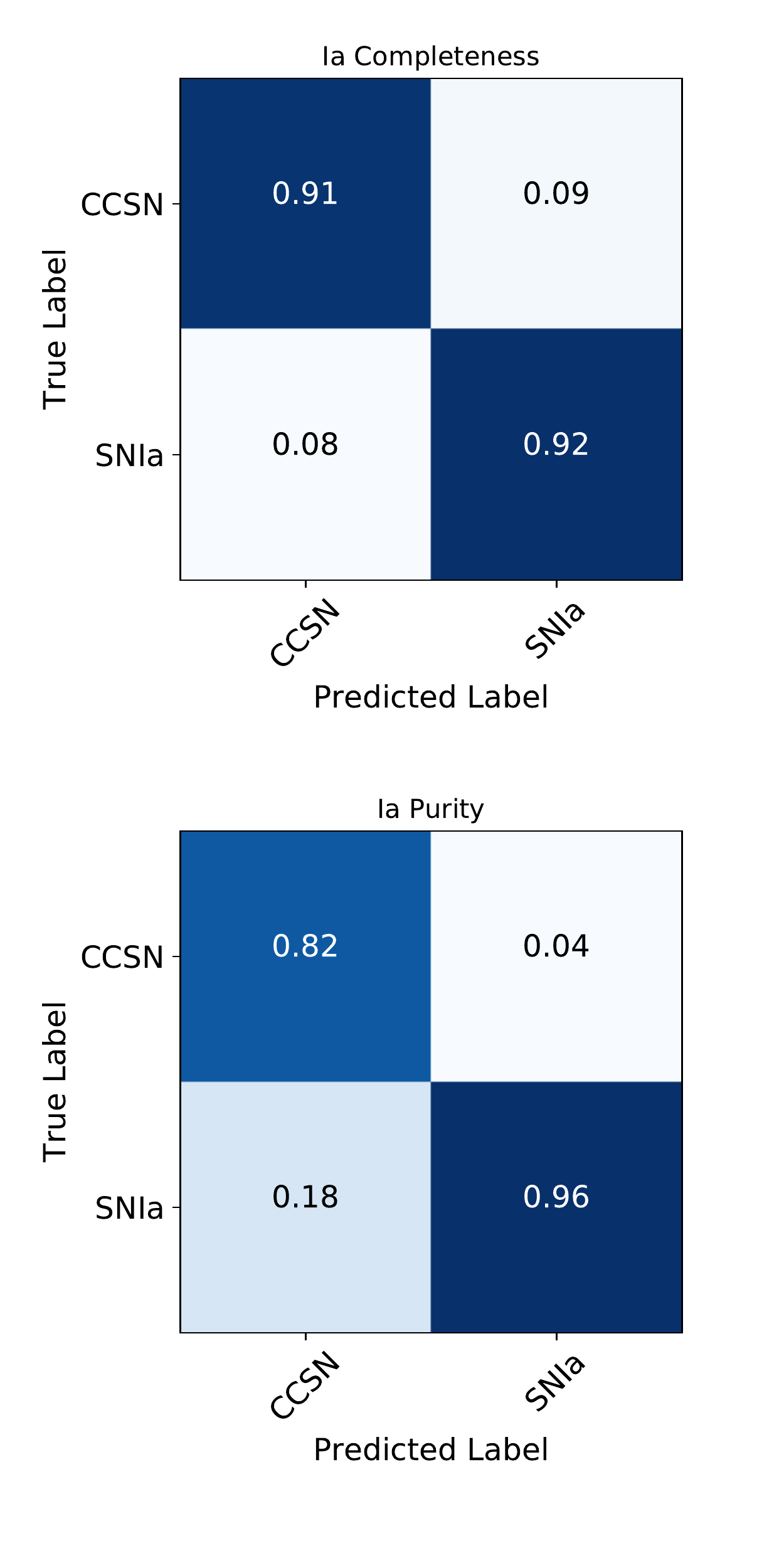}
\caption{\textit{Bottom:} Confusion matrices for a simpler Type Ia SN versus non-Ia (CCSN) classification, generated by collapsing the complete confusion matrices.  \label{fig:conf22}}
\end{figure} 

There are several metrics to measure the success of a classifier. We focus on three metrics: the purity, completeness and accuracy. We define the three, calculated for a single class, below:

\begin{equation}
    \begin{split}
        \mathrm{Purity} &= \frac{\mathrm{TP}}{\mathrm{TP}+\mathrm{FP}}\\
        \mathrm{Completeness} &= \frac{\mathrm{TP}}{\mathrm{TP}+\mathrm{FN}}\\
        \mathrm{Accuracy} &= \frac{\mathrm{TP}+\mathrm{TN}}{\mathrm{TS}}
    \end{split}
\end{equation}
where TP (FP) is the number of true (false) positives, TN (FN) is the number true (false) negatives, and TS is the total sample size. We optimize the hyperparameters of our classifier using the F1-score, defined here as the class-averaged harmonic mean of the purity and completeness.

\subsection{Spectroscopic Sample}

We visualize the completeness and purity of the spectroscopic sample using confusion matrices in Figure~\ref{fig:conf}.  A confusion matrix compares our RAENN label (horizontal axis, in the case of the completeness matrix) with the spectroscopic label (vertical). Results are shown for leave-one-out cross validation, in which one event is removed from the sample for training and the trained model is applied to the left out event. As with \citet{villar2019}, we find that our classifier performs best for Type Ia SNe (92\% completenss), SLSNe (76\%), and Type II SNe (82\%), and worst for Type Ibc SNe (37\%). Our class-averaged classification completeness is 69\% across the 5 SNe types. This is worse than the performance of \citet{villar2019}, who find a class-averaged completeness of 80\%. Our class-averaged purity is 66\%, again slightly worse than the average purity of 72\% found in \citet{villar2019}. When limiting the sample to only objects in which the classification probability is $\ge 0.7$ (a total of 438 objects), we find that our performance increases, with a class-averaged completeness of 75\% and a class-averaged purity of 74\% with a loss of  20\% of the sample size.

Next, we explore the classification confidence reported by our algorithm. The confidence estimates are directly outputted by the RF. With larger datasets, one can calibrate the outputted uncertainties using e.g., an additional logistic function. Given our small dataset, we do not perform any additional calibration. In Figure~\ref{fig:cumula} we plot the cumulative fractions of SNe in our training set, grouped by their spectroscopic and photometric classifications. The majority of high-confidence objects are Type Ia SNe, with nearly half of the spectroscopic Type Ia SNe having a confidence (p$>0.98$). Similarly, half of the SLSNe have high confidence identifications ($p>0.8$). Type Ibc SNe and Type IIn SNe have the lowest confidence on average, with the majority of events having $p<0.5$. This is likely reflective of the fact that Type Ibc and Type IIn SNe span a wide range of observed properties, including overlap with Type Ia SNe.

Figure~\ref{fig:cumula} also indicates the misclassified objects. Ideally, we want our misclassifications to largely occur in low-confidence objects. This is the case for SLSNe, Type Ia SNe and Type II SNe. For Type IIn and Type Ibc SNe, the misclassifications occur even for high-confidence events. This indicates that for Type Ia, Type II and SLSNe, misclassifications are likely tied to data quality. In contrast, misclassifications of Type IIn and Type Ibc SNe seem to be due to intrinsic overlap of the classes in feature-space with other SNe (mainly Type Ia SNe). 

We additionally attempt to sort events based on the number of data points, rather than classification confidence (see the right column of Figure~\ref{fig:cumula}). Our photometric dataset has, on average, fewer $>5\sigma$ datapoints compared to our spectroscopic dataset ($\sim15$ versus $\sim30$ data points on average). Because of this mismatch and the lack of a strong correlation between number of points and classification confidence, we do not further explore how cutting sparse light curves affects our final classification accuracy.

We next turn our attention to the performance of our classifier when constrained to only data-driven (RAENN) features. Using the same set of RAENN features \textit{without} any additional information, we produce the confusion matrices shown in Figure~\ref{fig:conf21}. We find a class-averaged completeness of 53\%, approximately 20\% worse than including the additional features. The overall breakdown is similar to our final confusion matrix, with the worst-performing classes being Type Ibc and Type IIn. We find that our RAENN-only classification is more inclined to label events as Type Ia SNe, likely a bias from the fact that our SN dataset used to train the RAENN is highly dominated by Type Ia SNe. If we run our classification algorithm \textit{without} the RAENN features, we find that {\tt SuperRAENN} performs similarly (slightly worse), implying that the RAENN has not picked up on uniquely helpful features independent from our hand-selected feature set. To be clear: the intent of RAENN is not necessarily to outperform hand-selected features but to create model-independent features in real time. In this work, we determine final classifications with the RAENN and hand-selected features to provide the highest confidence photometric classifications. Improvements to classifications based solely on RAENN features is left to future work.

While not optimized for Type Ia versus non-Type Ia SN classification, we explore how well our classifier (using the full set of features) performs when we collapse the confusion matrix into just two classes. In Figure~\ref{fig:conf22}, we show the completeness and purity confusion matrices for Type Ia versus non-Ia (CCSN) classifications, finding $\approx90$\% completeness and $>80$\% purity in both classes.

\begin{figure*}
\includegraphics[width=\textwidth]{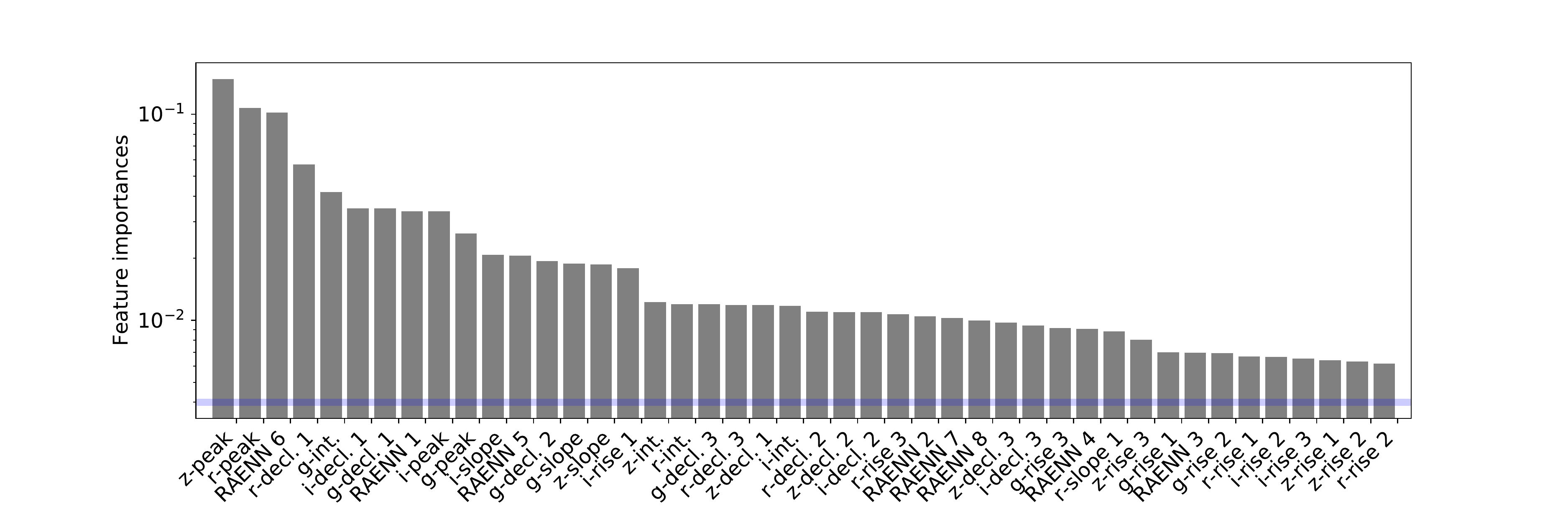}
\caption{Feature importance (grey). The blue horizontal line shows the importance measure for a normally-distributed random variable; features at or below this line can be considered largely unimportant to the final classification. In our case, all featured are considered important by the RF. \label{fig:featimport}}
\end{figure*}

The random forest classifier allows us to measure the relative ``importance" of the 44 features used to classify the SNe. We define importance as the decrease in the Gini impurity, which accounts for how often a feature is used to split a node and how often a node is reached in the forest \citep{breiman1984classification}.  We show the importance of each RF feature in Figure~\ref{fig:featimport}, along with the measured importance for a Normal random variable. The peak magnitudes and decline rates are the most important features for classification. However, the RAENN features also have significant influence on the final classifications, with two RAENN features appearing in the top ten important features. 

\begin{figure}
\includegraphics[width=0.5\textwidth]{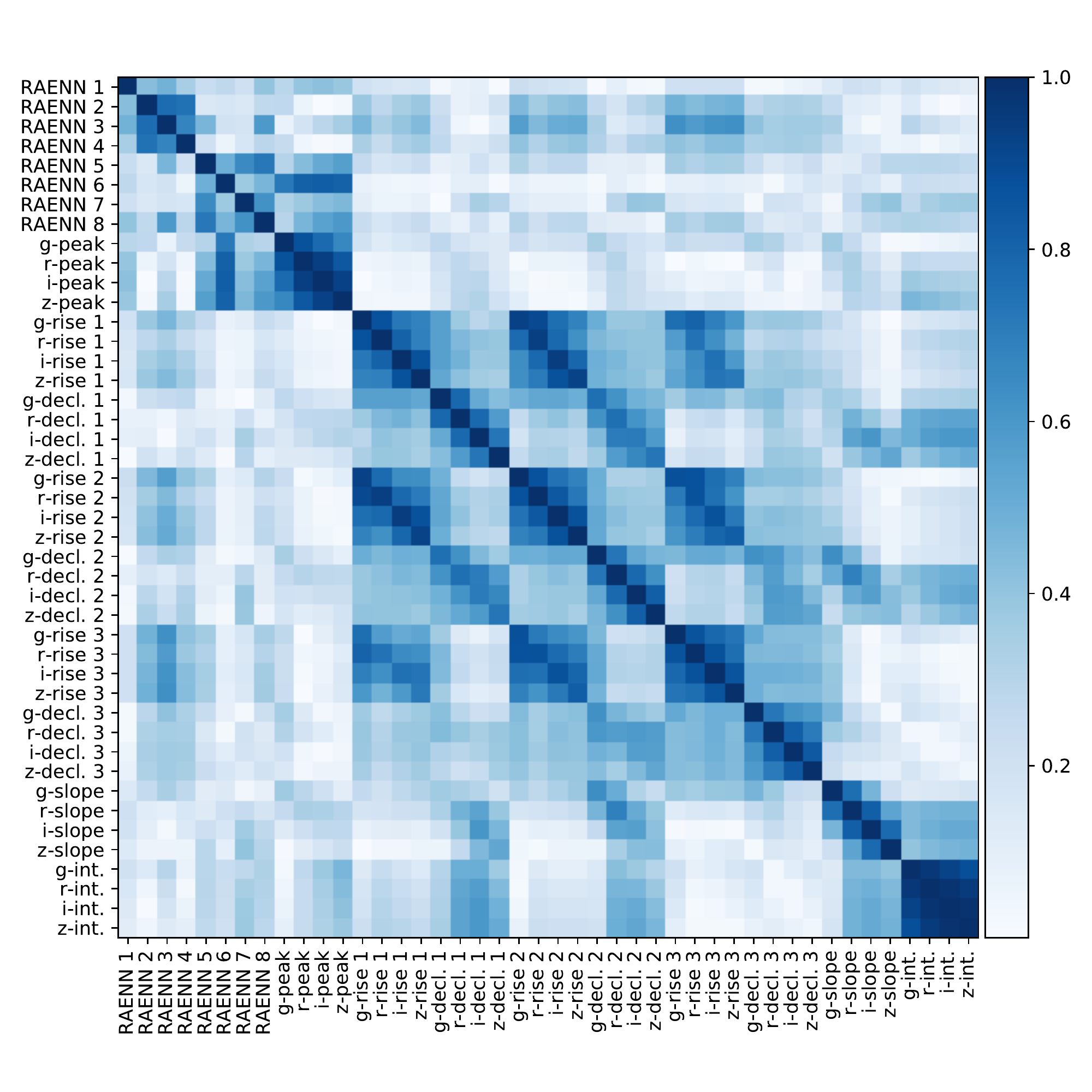}
\caption{Absolute values of the covariance matrix of the various features used in our classification method, where darker blue represents a stronger absolute correlation. Unsurprisingly, the same features derived from different bands (e.g., the peak $g$-band flux versus the peak $r$-band flux) are highly correlated. The RAENN features are also correlated to the physically-motivated parameters, with some being strongly correlated to peak magnitudes, some to rise and decline times, and some to the post-peak slope. \label{fig:dendro}}
\end{figure}

The feature importance unfortunately loses some quantitative meaning if the features are correlated, which is the case with our features. When two features are highly correlated, one may be arbitrarily measured as more important, so the general trends are more meaningful than precise order. We show the magnitudes of the feature correlations in Figure~\ref{fig:dendro} to better understand the underlying correlations. There are clear correlations between features derived in multiple bands (e.g., the peak magnitude in $g$-band is highly correlated to that in $r$-band). However, we also see correlations between the RAENN features and the more traditional light curve features. About half of the RAENN features seem strongly correlated with the peak magnitudes, while two others seem well-correlated with rise and decline times. A more detailed exploration of the physical interpretation of the RAENN feature-space may be worthwhile but is beyond the scope of this work.

\begin{figure*}
\centering
\includegraphics[width=0.8\textwidth]{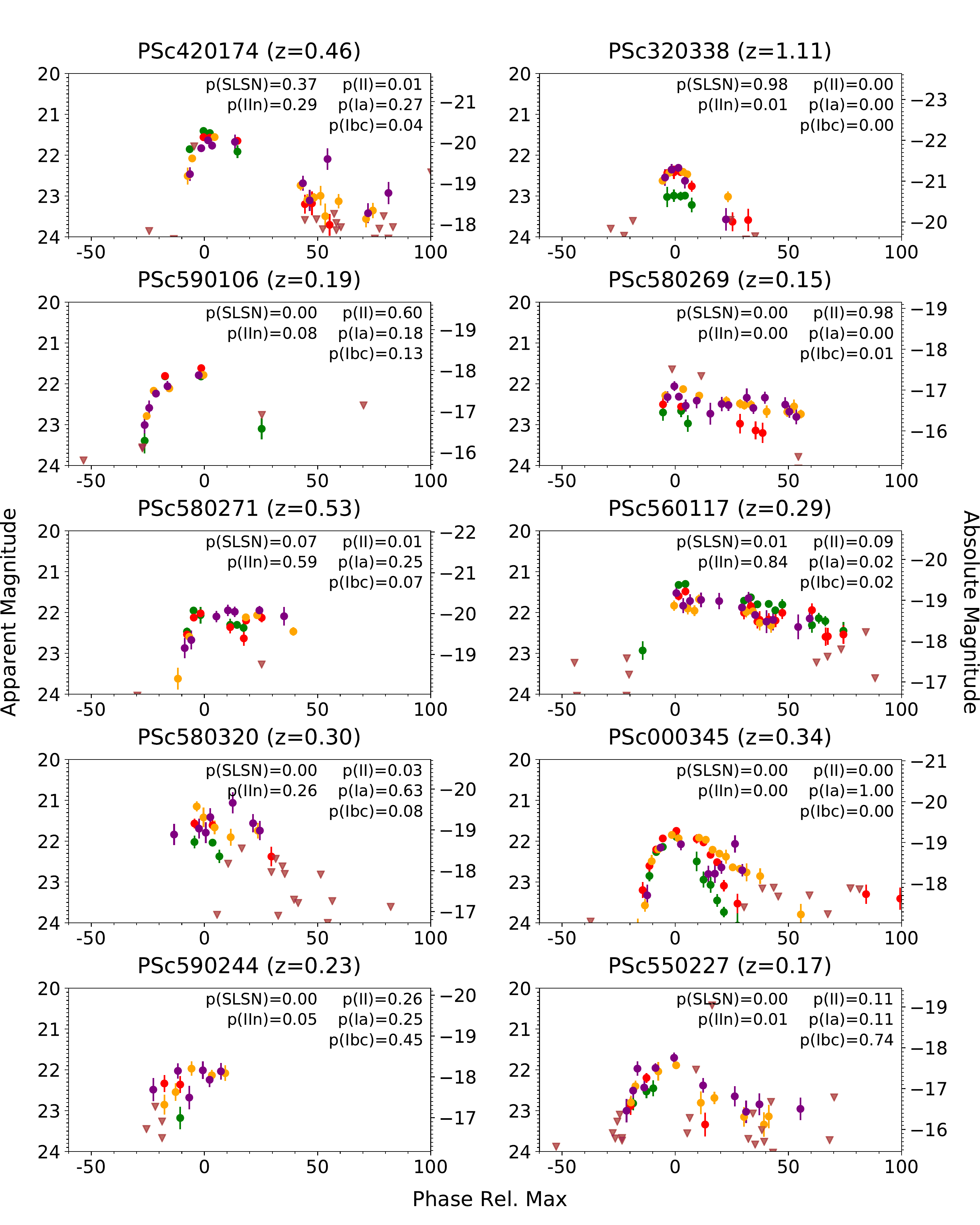}
\caption{A sample of SNe from our photometric sample, sorted by low (left column) versus high (right column) confidence and photometrically identified SN class (rows). Here we show only $>3\sigma$ detections and otherwise show magnitudes as upper limits (triangles). Low-confidence in classification appears to be both due either poor data quality or confusion between multiple classes.   \label{fig:samplelc}}
\end{figure*}

\subsection{Classifying the Complete Photometric Dataset}

\begin{figure*}
\includegraphics[width=\textwidth]{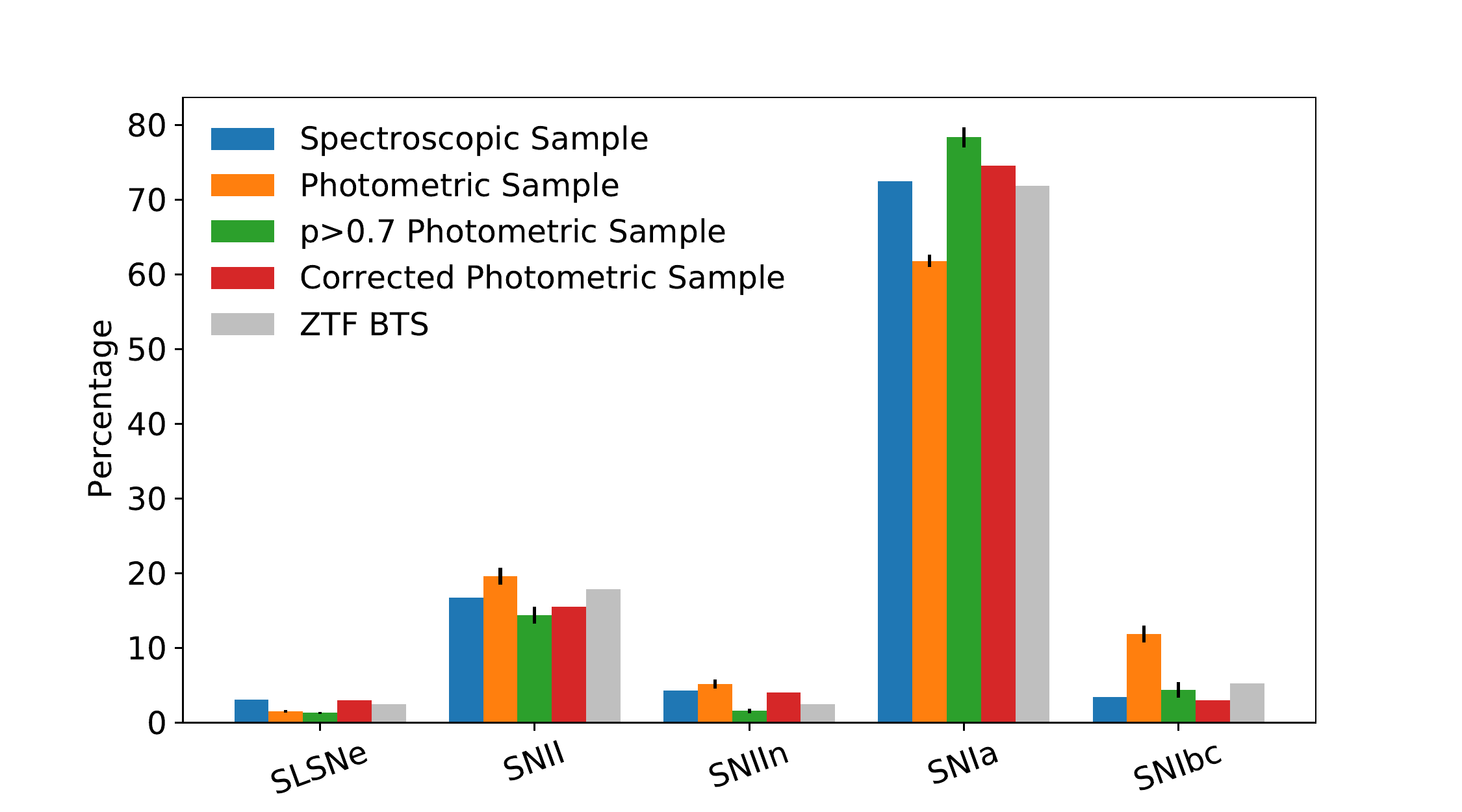}
\caption{Breakdown of SN subclasses in the spectroscopic and photometric samples. There is a significantly smaller fraction of Type Ia SNe in our photometric sample (orange) versus our spectroscopic sample (blue), implying we have misclassified Type Ia SNe as CCSNe. If we limit our photometric sample to the high-confidence ($p>0.7$) events (green), the class breakdowns are better aligned. Using our confusion matrix, we can correct the photometric class breakdown for known biases (see text for details; red), which also better aligns our class breakdowns. Finally, we compare our results to the ZTF Bright Transient Survey, finding good agreement between the spectroscopic class breakdown and corrected photometric class breakdown \citep{fremling2019zwicky}.\label{fig:pie}}
\end{figure*}

We apply our trained classification algorithm to the PS1-MDS dataset of SN-like transients that pass our quality cuts described in \S~\ref{sec:ps1}. We report the probabilities of each class type for each light curve in Table~\ref{tab:2}. Error bars for each class probability are calculated by running the trained RF classifier 25 times with unique random seeds. We show the class breakdown of the complete photometric set (\unsupsne\ SNe) in Figure~\ref{fig:pie}. Excluding the spectroscopic sample, we present \newsne\ new SNe with \newia\ (61.9\%) Type Ia SNe, \newiip\ (19.9\%) Type IIP SNe, \newibc\ (11.7\%) Type Ibc SNe, \newiin\ (4.8\%) Type IIn SNe,  and \newslsne\ (1.6\%) SLSNe. Of these, 1,311 are high-confidence ($p>0.7$) photometric classifications. A cumulative plot of the confidences grouped by each photometric class is shown in Figure~\ref{fig:cumula}; the distribution of these probabilities largely match the spectroscopic sample. 

A sample of SNe from each photometric class is shown in Figure~\ref{fig:samplelc}, including high- and low-probability examples. For the low-probability examples, it seems that even well-sampled light curves can have low confidence scores, likely because the features of their light curves reside on a region of feature space in which various SN classes reside. 

The redshift distributions of the new, photometrically-classified events is roughly consistent with that seen in the spectroscopic sample (Figure~\ref{fig:f1}), with Type Ibc peaking at $z\sim0.19$, Type II peaking at $z\sim0.21$, Type Ia and Type IIn peaking at $z\sim0.42$ and SLSNe peaking at $z\sim0.58$.

We compare the overall photometric breakdown of SN types to that of the ZTF Bright Transient Survey \citet{fremling2019zwicky}, which spectroscopically classified 761 SNe with peak $g$- or $r$-band magnitude of $<18.5$. \citet{fremling2019zwicky} find that their magnitude-limited survey breaks down into 72\% Type Ia SNe, 16\% ``normal" Type II SNe (Type IIP/L), 3\% Type IIn SNe (including their Type IIn and SLSN-II category), 5\% Type Ibc SNe, and 1.6\% Type I SLSNe. This is a similar breakdown found in our spectroscopic sample. Comparing to our photometric set, we find a slightly higher fraction of Type II and Type IIn SNe and a lower fraction of Type Ia SNe (all within $\sim20$\% of the ZTF BTS values), as shown in Figure \ref{fig:pie}. For our high-confidence ($p>0.7$) sample (also shown in Figure~\ref{fig:pie}), our class breakdowns are closer to those of our spectroscopic and the ZTF BTS sample, with a slight over-abundance of Type Ia SNe ($\approx78$\%). Based on our understanding of how our classifier performs on the training set, we can understand the biases present (e.g., that some spectroscopic Type Ibc SNe are classified photometrically as Type Ia SNe). We can use these known biases, encoded within the confusion matrices, to correct our class breakdown. Mathematically, this is calculated as the dot product of the purity matrix and our original class breakdown. Applying this correction to the photometric dataset, the class breakdown is well aligned with the breakdown of our spectroscopic sample, as shown in Figure~\ref{fig:pie}. This study should \textit{not} be used to rigorously study the observational breakdown of SN classes; however, the fact that our $p>0.7$ sample is in relatively good agreement with the ZTF BTS provides some evidence that our photometric sample is correctly labelled.

\section{Discussion}\label{sec:dis}

\subsection{Classification of Other Transients}\label{sec:dis1}
Our algorithm assumes that every SN belongs in one of five classes: SLSNe, Type II SNe, Type IIn SNe, Type Ia SNe and Type Ibc SNe. Yet what does our algorithm do for transients which do not fall in these five classes? Here we address this question for a number of spectroscopically classified extragalactic transients. We summarize the photometric classification for these rare transients in Table~\ref{tab:3}.

\citet{drout2014rapidly} presented a sample of ten extragalactic transients discovered with PS1-MDS with redshift measurements which rise too rapidly to be powered solely with $56$Ni\footnote{\citet{drout2014rapidly} presents an additional four objects which lack a confident redshift estimate (the ``bronze" sample), which we exclude from our analysis.}. Following \citet{rest2018fast}, we refer to these as FELTs. FELTs have a broad range of peak magnitude (-16 $\gtrsim M\gtrsim$ -20), which is reflected in the distribution of photometric classifications. Of these ten objects, six objects have ``high confidence" ($p>0.7$) classifications in one of our five categories: four of which are Type Ia SNe and two of which are Type II SNe. The other four objects are classified as low confidence Type Ia (one object), Type II (two objects) and Type Ibc (one object). As expected, the higher-luminosity objects are those classified as Type Ia, while the lower-luminosity objects are classified at Type II. The majority of objects have Type Ibc as their second-highest classification. Based on this analysis, FELTs are likely a (small) contaminant of both Type II and Type Ia SNe in our sample, and our algorithm would need to be retrained to specifically classify FELTs.

Two known TDEs were discovered in PS1-MDS: PS1-10jh (PSc040777, \citealt{gezari2012ultraviolet} and PS1-11af (PSc120170, \citealt{chornock2014tde}. Both objects are classified as Type IIn SNe with $p\sim0.8$ and $p\sim0.6$, respectively. This makes intuitive sense, as the light curves tend to be long-lived and bright like some Type IIn SNe. Both objects have Type Ia and Type II as their next most likely classifications. Based on these, it may be possible to search for TDEs in our sample within the photometric Type IIn sample.

We highlight four other SNe which do not fit in our five categories. PS1-10afx (PSc080333) is a lensed Type Ia SN \citep{chornock2013lensed,quimby2014lensed}, which peaks at -22 mag. We classify PS1-10afx as a high probability ($p\sim0.9$) SLSN. PS1-12sk (PSc370290) is a Type Ibn SN \citep{Sanders2013ibn} which peaks at $M\sim-19$. We classify PS1-12sk as a low probability Type Ia ($p\sim0.6$) or Type IIn ($p\sim0.4$). We classify PS1-12sz (PSc370330) as a likely IIb SN using SNID; PS1-12sz peaks at $M\sim-18.5$. We photometrically classify this object as a low probability Type Ibc ($p\sim0.6$). Finally, SN 2009ku (PS0910012) is a spectroscopically identified Type Iax \citep{narayan2011displaying} which peaks at $M\sim-18.5$. We classify this object as a low probability Type Ia ($p\sim0.5$) or Type Ibc ($p\sim0.3$).

\subsection{Potential Biases}

As discussed in \S~\ref{sec:ps1}, our spectroscopic sample is somewhat brighter and at a lower redshift than our test set. This difference may introduce biases in our final classifications, although this effect should be minimal considering the small ($\sim 1$ mag) difference between the two sets. De-redshifting the SNe removes some of this bias, by removing knowledge of the underlying redshift as a feature. 
The relative fractions of SN subtypes may evolve with redshift as host properties change (see e.g., \citealt{graur2017loss} for an exploration of the correlations between host properties and SN type). Our spectroscopic and photometric sets differs most greatly at $z\gtrsim 0.5$ (see Fig.~\ref{fig:f1}). In this redshift range, average host metallicity is not expected to drastically shift \citep{lilly2003metallicities}, implying a small potential bias. A separate bias may arise from the fact that our photometric sample relies on a measured spectroscopic redshift. At higher redshift, our galaxy redshift measurements become increasingly uncertain as dominant emission lines shift out of the optical band and intrinsically dim hosts fall below our observational limits. In contrast, rest-frame UV features of SNe (especially SLSNe) remain in the optical band, making it easier to confidently measure a distance from SNe spectra. In the future, this problem can be mitigated with photometrically derived host galaxy redshifts.

\begin{figure}
\includegraphics[width=0.5\textwidth]{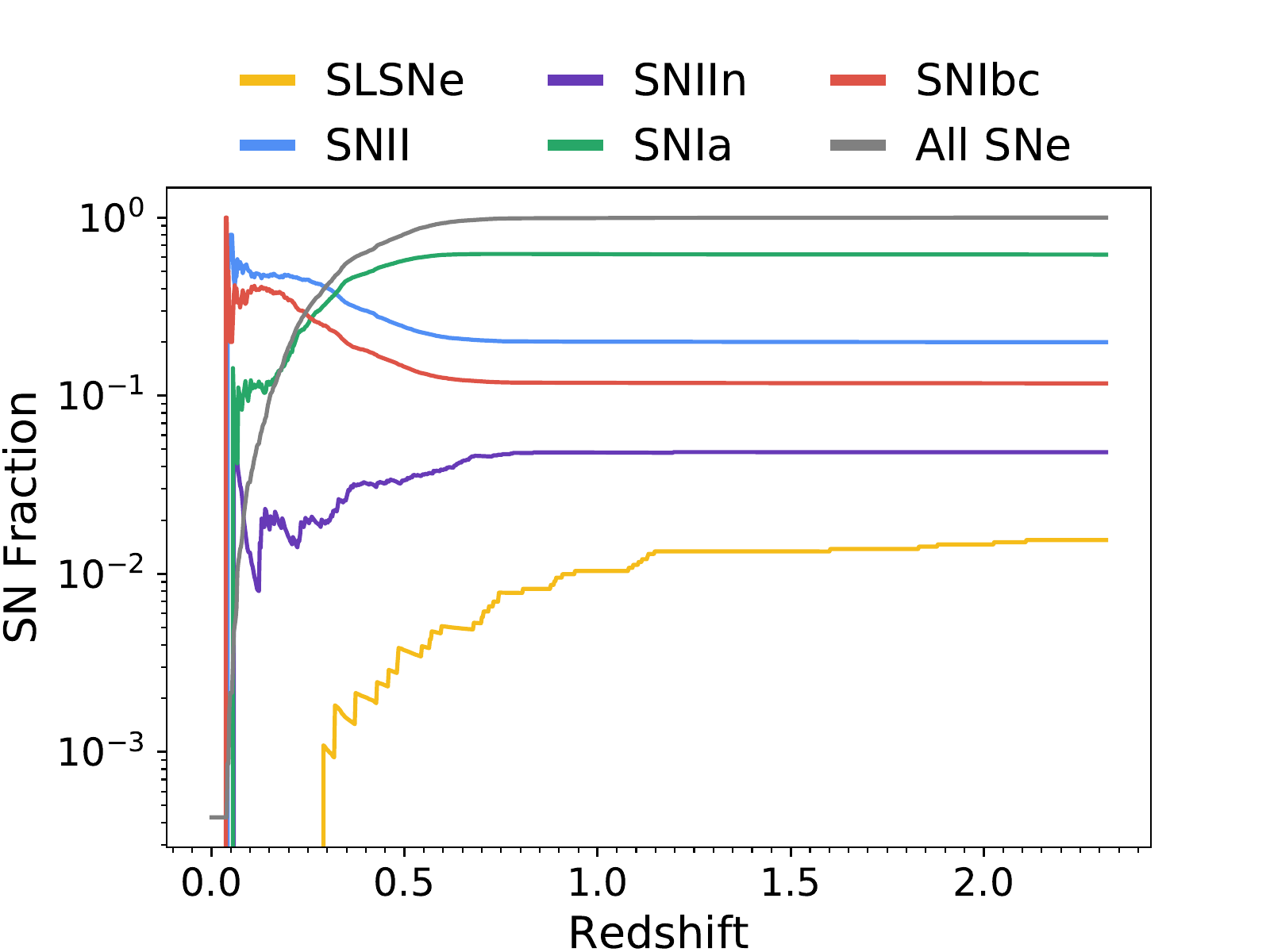}
\caption{Observed SN subclass cumulative fraction as a function of redshift (colored) and the overall cumulative distribution (grey).  \label{fig:zfrac}}
\end{figure}

As expected, the relative observed fraction of SN subtypes evolves with redshift due to the magnitude limit of the survey. We trace this evolution in Figure~\ref{fig:zfrac}. We show the cumulative fraction (integrating from $z=0$) of each subclass as a function of redshift. Each subclass peaks in order of luminosity function. The dimmest subclass, Type II SNe, dominates the sample for $z<0.3$, peaking near $z\sim 0$. 

Using the high redshift ($z>0.75$) sample, we can test if redshift information is playing an unwanted role in our training. The spectroscopic sample at $z\gtrsim0.75$ is solely made up of SLSNe; however, we do not expect \textit{all} high-$z$ objects to be SLSNe. Given a typical limiting magnitude of $m_\mathrm{r,lim}\sim23.3$, the corresponding absolute magnitude is $\sim-20$ at $z=0.75$. At this sensitivity, we expect to find SLSNe, Type IIn SNe and potentially bright Type Ia SNe (if the limiting magnitude is slightly deeper). For $z>0.75$, we find that our photometric sample (a total of 28 SNe) is 68\% SLSNe, 18\% Type IIn SNe and 14\% Type Ia SNe (with all Ia SNe occurring at $z<0.85$), implying our classifier has not learned to simply classify all high-$z$ events as SLSNe. The high-$z$ Type Ia SNe, in particular, have noisy light curves which peak at $M\sim-20$.

\subsection{Comparison to Other Works}

We first compare our results to \citetalias{hossen20}, which extends the work of \citet{villar2019} to classify the PS1-MDS photometric sample using features extracted from analytical fits to the light curves. Overall, \citetalias{hossen20} (and \citealt{villar2019}) achieve better performance at the cost of a more computationally-expensive feature extraction method. We agree with 74\% of the photometric classifications of \citetalias{hossen20}. If we compare the top two labels, the algorithms agree on 95\% of classifications. Indeed, often the top two classification choices are flipped for either algorithm, occurring most often with Type II/Ibc SNe and Type IIn/Ia SNe.  We find stronger agreement if we exclude objects with low classification confidence; namely, using only $p>0.7$ in both algorithms, our classifications agree 84\% of the time (with 1,597 objects remaining after the cut, i.e., a loss of $\sim50$\% of the sample). The agreement increases further for even higher probability cuts of $p>0.8$ ($>0.9$), with 88\% (92\%) agreement with 1249 (888) objects remaining. Most classification disagreements lie in the Type Ibc/IIn categories, which have low confidence classifications. We find that our algorithm is more likely to classify SNe as Type Ia, likely a bias built into the unsupervised step of training on the complete dataset (which is dominated by Type Ia SNe). A more detailed comparison of these two results is provided in \citetalias{hossen20}.

\citet{villar2019} discusses the difficulty in comparing our results to the broader literature. In short, previous works have largely focused on Type Ia versus CCSN classification (e.g., \citealt{ishida2013kernel,jones2017measuring,brunel2019cnn}) or have been trained and tested on simulated data (e.g., \citealt{kessler2010results,muthukrishna2019rapid,moller2020supernnova}). In the case of Type Ia versus CCSN classification, we achieve an accuracy of $\approx92$\%, similar to (but somewhat worse than) specialized classifiers \citep{jones2017measuring,brunel2019cnn}. 
When comparing to works based on simulated data, we caution that not all simulated datasets are suitable for multi-class SN classification. In particular, the Supernova Photometric Classification Challenge (SNPCC) training set \citep{kessler2010results} lacks the SN diversity necessary to accurately train classifiers and will lead to artificially promising results. PLAsTiCC \citep{allam2018photometric,kessler2019models} is better suited for this task, and we encourage future work to be built on this dataset or the PS1-MDS dataset presented here. 

We next compare our results to \citealt{jones2017measuring}, who presented a PS1-MDS sample of 1,169 likely Type Ia SNe, focusing on Type Ia versus non-Ia classification. \citealt{jones2017measuring} used four classification algorithms: the template-matching algorithm PSNID \citep{sako2011photometric}, a nearest neighbor approach using the PSNID templates; an algorithm based on fitting light curves to SALT2 templates \citep{guy2007salt2}; and a method, GALSNID \citep{foley2013classifying}, which only utilizes host galaxy properties. \citet{jones2017measuring} similarly removed objects with unreliable host redshifts and potential AGN hosts, but unlike our analysis they removed objects at $z>0.75$. Of their 1,169 identified Type Ia SNe, only 1,046 SNe pass our quality cuts to be classified in this work. For these, we find 95\% agreement. Of the remaining 48 SNe, we identified Type Ia as the second highest choice in 24 cases. Of the remaining 24 cases, 15 have low Type Ia probabilities ($p<0.8$ from \citealt{jones2017measuring}) or classification probabilities based entirely on host galaxy. It is worth noting that our classifier, similar to \citet{jones2017measuring}, achieves 96\% purity in Type Ia SNe, making it likely usable for cosmological studies \citep{jones2018measuring}. 

We compare our results to those trained on PLAsTiCC -- in particular, \citealt{boone2019,muthukrishna2019rapid} and \citealt{Gabruseva2020}. These classifiers present average completenesses of $\approx0.88$ for SLSNe (higher than our score), $\approx0.5$ for Type II/IIn SNe (lower than our averaged Type II/IIn score), $\approx0.92$ for Type Ia SNe (similar to our score), and $\approx0.46$ for Type Ibc SNe (similar to our score given low-number statistics). These results are based on simulated data which lack the complexity of real data, so it is encouraging that our algorithm performs similarly or outperforms these works. It would be interesting and useful to the community to know how these algorithms perform on the PS1-MDS dataset, but we leave this for future work.

\begin{figure*}
\hspace*{-2cm}\includegraphics[width=1.25\textwidth]{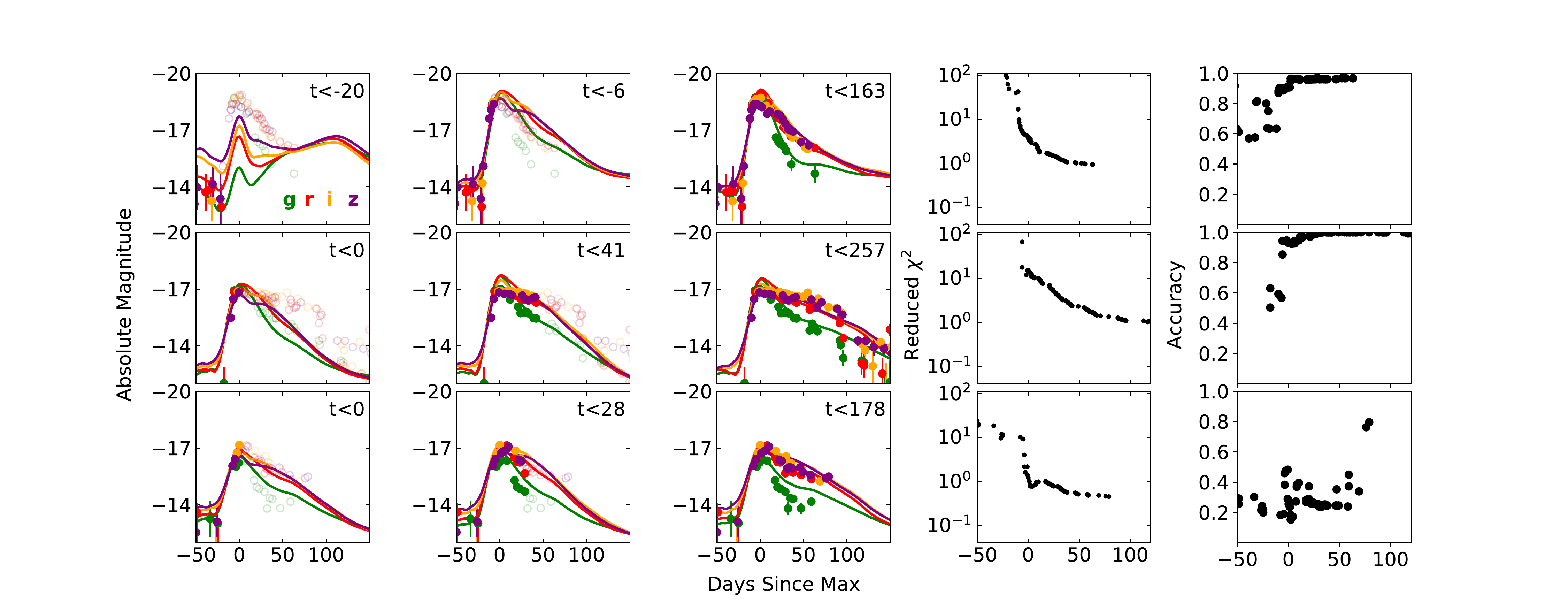}
\caption{Examples of a Type Ia SN (\textit{top row}), Type II SN (\textit{middle row}) and Type Ibc SN (\textit{bottom row}). Filled points represent observations used to generate the RAENN model (colored lines), while empty points are the full data set to guide the eye. In the right-most column, we show the root-mean-squared (RMS) error as a function of SN phase, as more data are being included in the RAENN model. Interestingly, the RMS reaches $\sim1$ near peak for all SNe shown. We emphasize that the RAENN model has been optimized to classify complete SN light curves rather than partial light curves.\label{fig:extrap}}
\end{figure*}

\subsection{RAENN Architecture: Limitations and Benefits}
\begin{figure}[t]
\includegraphics[width=0.5\textwidth]{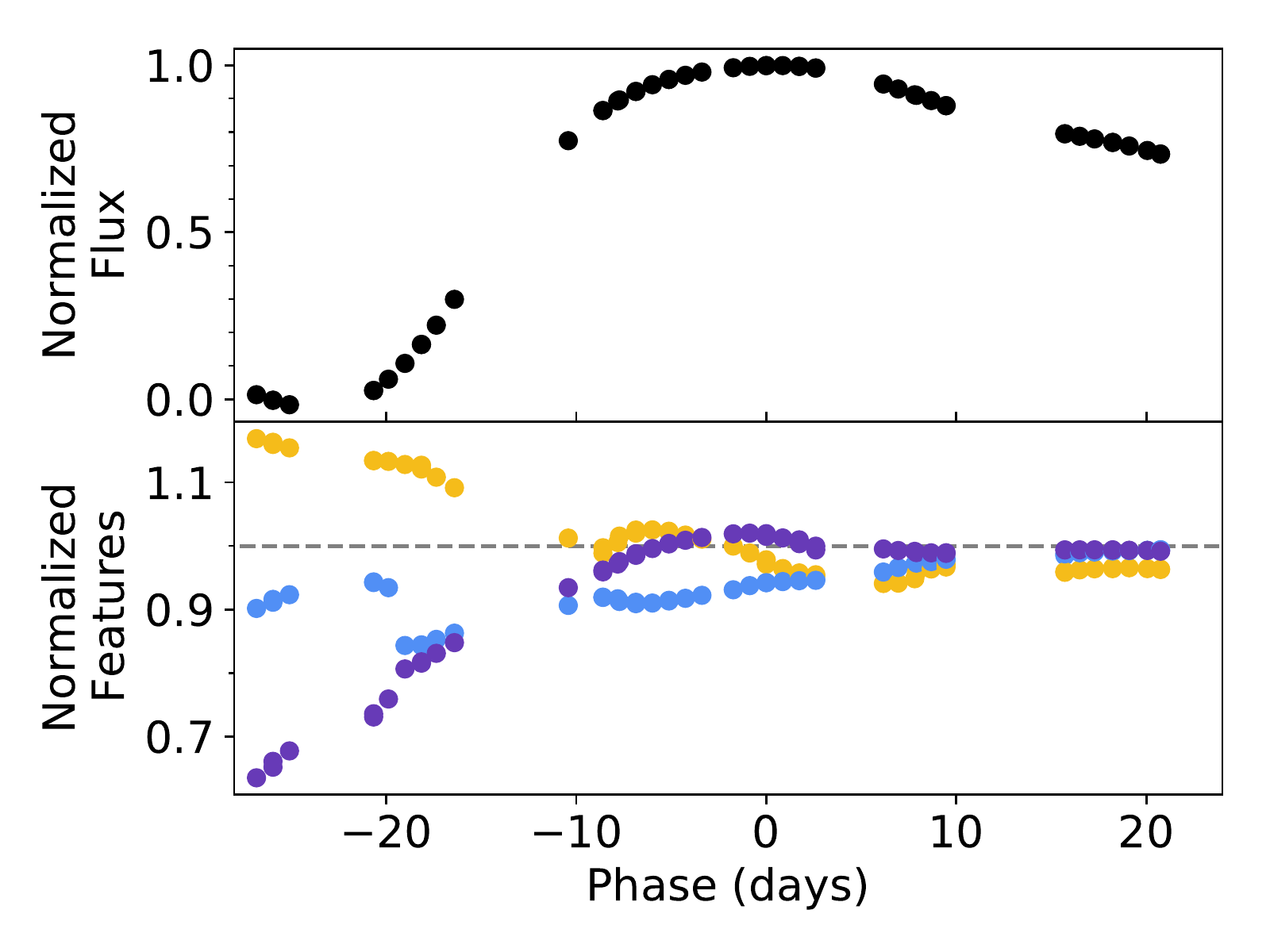}
\caption{\textit{Top:} Normalized, GP-interpolated $r$-band light curve of a spectroscopically-classified Type Ia SN. \textit{Bottom:} Representative set of three (orange, purple, blue) normalized AE features as a function of SN phase. To generate these features, we run the light curve data through the RAENN up to a certain phase. As shown, the values vary smoothly and then settle on the final values about one week post-peak. \label{fig:en1}}
\end{figure}

We now turn to the architecture of the RAENN itself and its use in future surveys. The recurrent neurons allow our neural network to generate light curve features that can be updated in real time, in addition to extrapolating and interpolating light curves. We highlight the accuracy of the RAENN light curve model as a function of light curve completeness in Figure~\ref{fig:extrap}. We track how well the RAENN is able to both model the complete light curve and accurately classify the SNe with limited data by providing a partial light curve into the RAENN. For each step, we hold the other features (e.g., peak luminosity and duration) constant. This is not a completely robust method, as some features (e.g., decline time) cannot be measured before peak. We leave the optimization of {\tt SuperRAENN} for real time data streams to future work. We find that {\tt SuperRAENN} performance  drastically improves post-peak, but that it can provide accurate classifications and light curves somewhat before peak. To explore why {\tt SuperRAENN} improves near-peak, we track how the RAENN features change as the light curves evolve. In Figure~\ref{fig:en1}, we plot the values of representative encoding values of a Type Ia SN. The encodings vary smoothly until settling on the correct final values $\sim10$ days post-peak. 

\begin{figure}[b]
\includegraphics[width=0.5\textwidth]{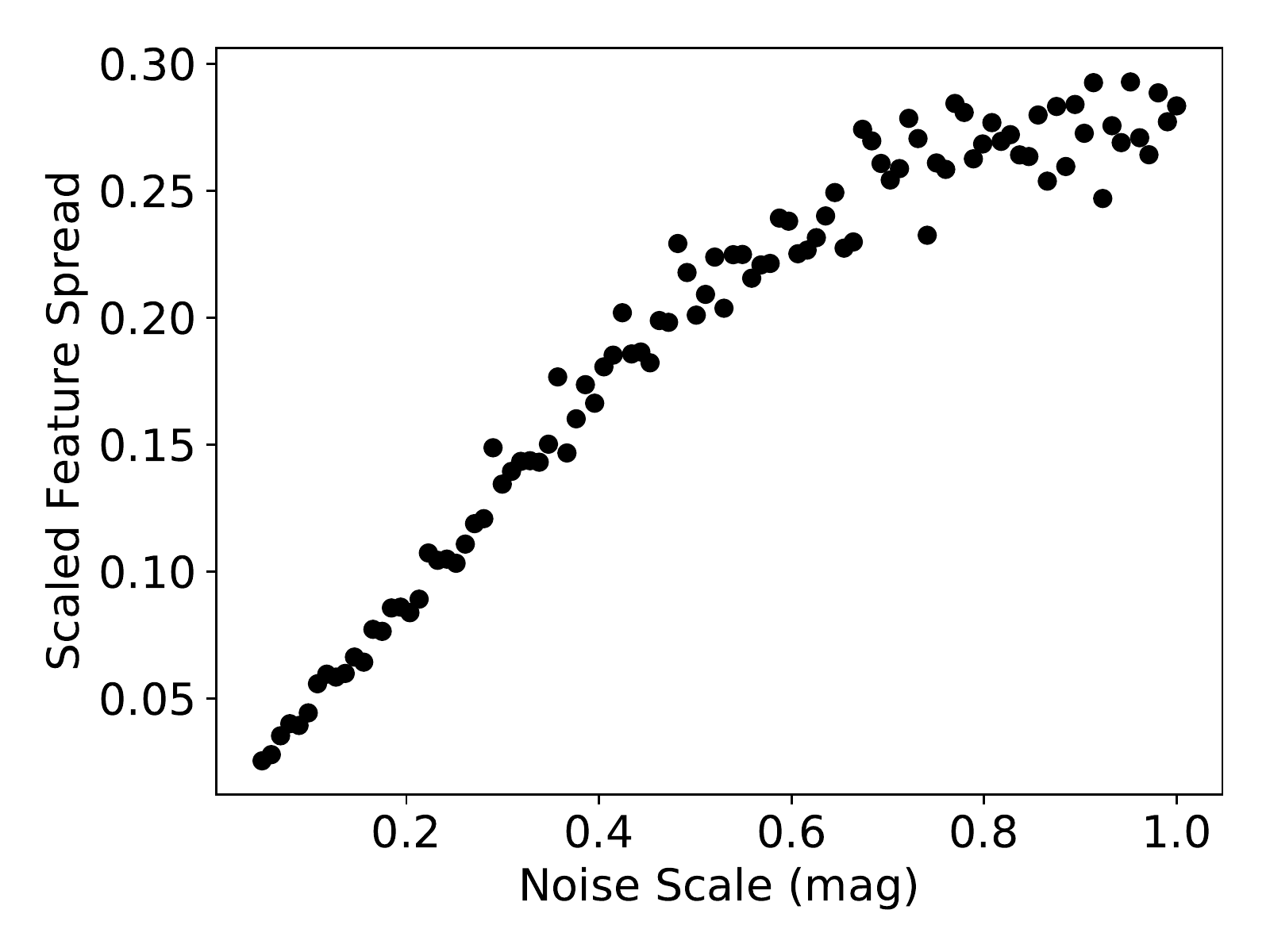}
\caption{Average spread of the RAENN features for a spectroscopic Type Ia SN as a function of light curve noise. For every noise scale, we run 100 simulations, adding random noise to the light curve. We then track the average spread of each parameter. We scale this spread by the total spread in the Type Ia class. Even with an injected error of 0.5 mag, the spread in the RAENN feature space only reaches 30\% of the total spread throughout the Type Ia class in feature space, implying the method is robust to noise. \label{fig:en2}}
\end{figure}

The ability of the RAENN to extrapolate light curves without built-in physical assumptions allows it to search for anomalous events in real time for the purpose of spectroscopic and multi-wavelength follow-up. Given the millions of events expected from LSST, it is essential to search for unexpected or previously unknown physical effects that. One concern is that our algorithm is potentially not robust to noisy live-streaming data; in other words, our algorithm must be able to distinguish between anomalous data and noisy data. We check the stability of our encoded values as a function of scaled white noise by adding white noise to a light curve. We then use our RAENN to encode the noisy light curve and record the scatter of the encoded values. We report the results of this test in Figure~\ref{fig:en2}, in which we show the scaled scatter of the encoded values as a function of the magnitude of the injected noise. The scatter grows linearly with noise; however, even with one magnitude of scatter added to the light curve, the overall scatter of the encoded values only increases to 30\% of the overall spread of class's features. This implies that the RAENN is largely robust to noise.

Several steps need to be taken to allow our architecture to work on streaming data. First, we use phases relative to maximum light, which will be unavailable during the rise of the SN. A shift to a time measurable early in the light curve, like time of first detection, will allow the RAENN to otherwise perform as designed. Similarly, the features utilized during the supervised portion of our classifier rely on the full light curve being available. All features can be estimated from extrapolated RAENN light curves or a new set of features may be used on streaming data. Finally, although not necessary, our RAENN could output uncertainties on the SN light curves by converting the network into a \textit{variational} AE, which is designed to simultaneously find an encoding space and uncertainties on the encoded data. This more complex architecture would likely require a larger training set to be reliable. Finally, we note that an algorithm like RAENN could be used in conjunction with an active oracle (a software which recommends new observations to improve classification) such as REFITT \citep{2020Sravan}, in order to actively optimize classification accuracy.

\section{Conclusions}\label{sec:con}

Deep learning-based classifiers are becoming increasingly important for classification of archival SN light curves. In this paper we present a novel, semi-supervised approach to light curve classification, which utilizes spectroscopically labelled and unlabelled SN data from the PS1-MDS. Our key conclusions are as follows:

\begin{enumerate}
    \item We present the light curves of \allsne\ SN-like events discovered with PS1-MDS. 
    \item We present the spectroscopic classifications of \specsne\ SNe, including \numslsne\ Type I SLSNe, \numiip\ Type II, \numiin\ Type IIn, \numia\ Type Ia and \numibc\ Type Ibc SNe.
    \item We measure and report the spectroscopic redshifts for \unsupsne\ SN-like events used in our unsupervised training set.
    \item We present a new, open source photometric classification algorithm, {\tt SuperRAENN}. {\tt SuperRAENN} uses a semi-supervised approached and novel neural network architecture to classify irregularly-sampled SN light curves. 
    \item Using {\tt SuperRAENN}, we extract learned, nonlinear features from the sparse light curves. We use these features and others to classify the complete set of \unsupsne\ SN-like objects in the PS1-MDS dataset with host galaxy redshifts.
    \item We achieve high (87\%) accuracy for our spectroscopically labelled sample. We find best performance for SLSNe, Type Ia, and Type II SNe due to their distinctive regions of feature space. We find worst performance for Type Ibc SNe, likely due to the small sample size (just 19 events) and their significant overlap with Type Ia SNe and the subset of rapidly-declining Type II SNe (formerly, IIL).
    \item Compared to previous studies, we find that our general classifier performs as well or can outperform classifiers trained on synthetic data sets.
    \item  We perform simple tests for classification bias and method robustness to noise, finding our method robust to both.
\end{enumerate}

In addition to these key results, we highlight several lessons learned from this study. We find that both Type IIn and Type Ibc classes suffer from poor accuracy likely due to substantial overlap with Type Ia SNe in feature space. This finding has also been shown in \citet{villar2019} and \citetalias{hossen20}, implying this is a general problem for classifiers. Additionally, rare transients, e.g. FELTs, abnormal Type Ia classes, etc., can be hidden as high-confidence events in another class \textit{or} low-confidence events across several classes. Adapting pre-existing classifiers to new classes should be taken on a case-by-case basis. Finally, we find that a mixture of hand-selected and data-driven (in our case, RAENN) features can improve classification accuracy, but hand-selected features seem to generally out perform data-driven features.

Finally, we note that several modifications to our presented classifier will allow it to work with live, rather than archival, data streams such as ZTF and LSST. We perform simple tests and find that our classifier performs optimally around peak, although we have not optimized for this purpose. Finally, the RAENN architecture may also be utilized to search for anomalous events in real time. We plan to explore this in future work.

\newpage
\begin{deluxetable*}{cccccccccccc p{2cm} cc}
\rotate
\tabletypesize{\scriptsize}
\caption{SNe Properties\label{tab:one}}
\tablehead{\colhead{Object} & \colhead{PSc ID}  & \colhead{IAU Name} & \colhead{CBET} & \colhead{SN RA} & \colhead{SN Dec} & \colhead{Milky Way} & \colhead{SN Type} & \colhead{Redshift$^a$} & \colhead{Host RA} & \colhead{Host Dec} & \colhead{Num.} & \colhead{Telescope$^b$/} & \colhead{Unsup.$^d$} & \colhead{Sup.$^e$} \vspace{-0.15cm}\\
\colhead{Name} & \colhead{}  & \colhead{} & \colhead{} & \colhead{(deg.)} & \colhead{(deg.)} & \colhead{E(B-V)} & \colhead{} & \colhead{} & \colhead{(deg.)} & \colhead{(deg.)} & \colhead{Points$^c$} & \colhead{Source} & \colhead{} & \colhead{} }
\startdata
PS1-0909006 & PS0909006 & SN 2009ks & \cite{2009CBET.2012....1R} & 333.9503 & 1.1848 & 0.0426 & SNIa & 0.284 & - & - & 19 & Gem-N & Y & Y \\
PS1-0909010 & PS0909010 & SN 2009kt & \cite{2009CBET.2012....1R} & 37.1182 & -4.0789 & 0.0256 & SNIa & 0.27 & - & - & 20 & Gem-N & Y & Y \\
PS1-0910012 & PS0910012 & SN 2009ku & \cite{2009CBET.2012....1R} & 52.4718 & -28.0867 & 0.0073 & SNIax & 0.079 & - & - & 40 & Gem-S, NOT, Magellan, Gem-S & Y & N \\
PS1-0910016 & PS0910016 & SN 2009kx & \cite{2009CBET.2012....1R} & 35.3073 & -3.91 & 0.0219 & SNIa & 0.23 & - & - & 10 & Gem-N & Y & Y \\
PS1-0910017 & PS0910017 & SN 2009kv & \cite{2009CBET.2012....1R} & 35.2775 & -5.0233 & 0.0221 & SNIa & 0.32 & - & - & 15 & Gem-N & Y & Y \\
PS1-0910018 & PS0910018 & SN 2009kz & \cite{2009CBET.2012....1R} & 35.667 & -4.0273 & 0.0242 & SNIa & 0.265 & - & - & 11 & Gem-N & Y & Y \\
PS1-0910020 & PS0910020 & SN 2009kw & \cite{2009CBET.2012....1R} & 54.5975 & -28.2533 & 0.013 & SNIa & 0.242 & - & - & 13 & Gem-S & Y & Y \\
PS1-0910021 & PS0910021 & SN 2009ky & \cite{2009CBET.2012....1R} & 53.62 & -27.9084 & 0.0081 & SNIa & 0.256 & - & - & 20 & Gem-S & Y & Y \\
PS1-10a & PSc000001 & - & - & 52.4531 & -29.075 & 0.009 & SNII & 0.071 & 52.4536 & -29.0744 & 13 & 2df & Y & Y \\
PS1-10aa & PSc010046 & - & - & 162.9188 & 58.1822 & 0.0115 & - & 0.039 & 162.9188 & 58.1822 & 21 & WIYN & N & N \\
PS1-10aaa & PSc070003 & - & - & 333.0415 & 0.7193 & 0.0479 & - & 0.1376 & 333.0414 & 0.7194 & 63 & NED, SDSS & Y & N \\
PS1-10aab & PSc070004 & - & - & 335.0863 & 0.524 & 0.0641 & - & - & 335.088 & 0.5244 & 31 & - & N & N \\
PS1-10aac & PSc070039 & - & - & 334.6541 & 0.6184 & 0.0587 & - & 0.3933 & 334.6538 & 0.6186 & 28 & MMT & N & N \\
PS1-10aad & PSc070048 & - & - & 241.0171 & 54.1988 & 0.0088 & - & 0.423 & 241.017 & 54.1988 & 15 & MMT & Y & N \\
\enddata
\tablecomments{A complete, machine-readable version of this table is available in the online version.\\
a. Redshift estimate from either the transient spectra or host galaxy spectra.\\
b. Telescope used to acquire galaxy or SN spectra. \\
c. Number of $>5\sigma$ datapoints in light curve. \\
d. Included in unsupervised training set. These objects have reliable host redshift estimates. \\
e. Included in supervised training set.}
\end{deluxetable*}

\begin{deluxetable*}{cccccc}
\caption{SNe classification\label{tab:2}}
\tablehead{\colhead{Event Name} & \colhead{p$_\mathrm{SLSN}$} & \colhead{p$_\mathrm{II}$} & \colhead{p$_\mathrm{IIn}$} & \colhead{p$_\mathrm{Ia}$} & \colhead{p$_\mathrm{Ibc}$}}
\startdata
PSc000001 & $0.00_{0.00}^{0.00}$ & $1.00_{0.00}^{0.00}$ & $0.00_{0.00}^{0.00}$ & $0.00_{0.00}^{0.00}$ & $0.00_{0.00}^{0.00}$\\
PSc000006 & $0.00_{0.00}^{0.00}$ & $0.00_{0.00}^{0.00}$ & $0.00_{0.00}^{0.00}$ & $1.00_{0.00}^{0.00}$ & $0.00_{0.00}^{0.00}$\\
PSc000010 & $0.00_{0.00}^{0.00}$ & $0.00_{0.00}^{0.00}$ & $0.00_{0.00}^{0.00}$ & $1.00_{0.00}^{0.00}$ & $0.00_{0.00}^{0.00}$\\
PSc000011 & $0.00_{0.00}^{0.00}$ & $0.00_{0.00}^{0.00}$ & $0.00_{0.00}^{0.00}$ & $1.00_{0.00}^{0.00}$ & $0.00_{0.00}^{0.00}$\\
PSc000014 & $0.00_{0.00}^{0.00}$ & $0.00_{0.00}^{0.00}$ & $0.00_{0.00}^{0.00}$ & $1.00_{0.00}^{0.00}$ & $0.00_{0.00}^{0.00}$\\
PSc000034 & $0.00_{0.00}^{0.00}$ & $0.00_{0.00}^{0.00}$ & $0.00_{0.00}^{0.00}$ & $1.00_{0.00}^{0.00}$ & $0.00_{0.00}^{0.00}$\\
\enddata
\tablecomments{A complete, machine-readable version of this table is available in the online version. Spectroscopically identified SNe have probabilities of one.}
\end{deluxetable*}

\begin{deluxetable*}{cccccccc}
\caption{Rare Transients Classification\label{tab:3}}
\tablehead{\colhead{Event Name} & \colhead{Spec. Class} & \colhead{Ref.} & \colhead{p$_\mathrm{SLSN}$} & \colhead{p$_\mathrm{II}$} & \colhead{p$_\mathrm{IIn}$} & \colhead{p$_\mathrm{Ia}$} & \colhead{p$_\mathrm{Ibc}$}}
\startdata
PSc040777 (PS1-10jh) & TDE & \cite{gezari2012ultraviolet} & 0.06 & 0.06 & \textbf{0.79} & 0.06 & 0.03 \\
PSc120170 (PS1-11af) & TDE & \cite{chornock2014tde} & 0.12 & 0.08 & \textbf{0.62} & 0.14 & 0.04 \\
PSc080333 (PS1-10afx) & Lensed Ia & \cite{chornock2013lensed} & \textbf{0.88} & 0.00 & 0.09 & 0.02 & 0.01 \\
PSc370290 (PS1-12sk) & Ibn & \cite{Sanders2013ibn} & 0.00 & 0.00 & 0.41 & \textbf{0.57} & 0.02\\
PS0910012 (SN 2009ku) & Iax & \cite{narayan2011displaying} & 0.00 & 0.13 & 0.04 & \textbf{0.55} & 0.28 \\
PSc010411 (PS1-10ah) & FELT & \cite{drout2014rapidly} & 0.00 & \textbf{0.74} & 0.01 & 0.12 & 0.13 \\
PSc091902 (PS1-10bjp) & FELT & \cite{drout2014rapidly} & 0.00 & \textbf{0.56} & 0.27 & 0.14 & 0.03 \\
PSc150020 (PS1-11qr) & FELT & \cite{drout2014rapidly} & 0.00 & 0.02 & 0.06 & \textbf{0.82} & 0.09 \\
PSc340012 (PS1-11bbq) & FELT & \cite{drout2014rapidly} & 0.03 & 0.00 & 0.04 & \textbf{0.80} & 0.13 \\
PSc350224 (PS1-12bb) & FELT & \cite{drout2014rapidly} & 0.00 & 0.28 & 0.13 & 0.17 & \textbf{0.42} \\
PSc350352 (PS1-12bv) & FELT & \cite{drout2014rapidly} & 0.00 & 0.00 & 0.00 & \textbf{0.95} & 0.04 \\
PSc440088 (PS1-12brf) & FELT & \cite{drout2014rapidly} & 0.00 & 0.14 & 0.11 & \textbf{0.49} & 0.25 \\
PSc570006 (PS1-13duy) & FELT & \cite{drout2014rapidly} & 0.00 & 0.02 & 0.08 & \textbf{0.79} & 0.12 \\
PSc570060 (PS1-13dwm) & FELT & \cite{drout2014rapidly} & 0.00 & \textbf{0.76} & 0.03 & 0.09 & 0.13 \\
PSc580304 (PS1-13ess) & FELT & \cite{drout2014rapidly} & 0.00 & \textbf{0.51} & 0.02 & 0.37 & 0.11 \\
\enddata
\end{deluxetable*}

\facilities{ADS, NED, PS1, TNS}
\defcitealias{astropy_collaboration_astropy_2018}{Astropy Collaboration 2018}
\software{Astropy \citepalias{astropy_collaboration_astropy_2018}, extinction \citep{barbary_extinction_2016}, keras \citep{chollet2015keras}, Matplotlib \citep{hunter_matplotlib:_2007}, NumPy \citep{oliphant_guide_2006}, RVSAO \citep{kurtz_rvsao_1998}, scikit-learn \citep{pedregosa_scikit-learn_2011}, SciPy \citep{virtanen_scipy_2020}}

\acknowledgements
We thank Jessica Mink and Brian Hsu for providing assistance with host galaxy redshifts estimates. V.A.V.~acknowledges support by the Ford Foundation through a Dissertation Fellowship and the Simons Foundation through a Simons Junior Fellowship (\#718240). G.H. thanks the LSSTC Data Science Fellowship Program, which is funded by LSSTC, NSF Cybertraining Grant \#1829740, the Brinson Foundation, and the Moore Foundation; his participation in the program has benefited this work. The Berger Time-Domain Group is supported in part by NSF grant AST-1714498 and the Harvard Data Science Initiative. D.O.J. is supported by a Gordon and Betty Moore Foundation postdoctoral fellowship at the University of California, Santa Cruz. R.L. is supported by a Marie Sk\l{}odowska-Curie Individual Fellowship within the Horizon 2020 European Union (EU) Framework Programme for Research and Innovation (H2020-MSCA-IF-2017-794467). D.M. acknowledges NSF support from from grants PHY-1914448 and AST-2037297. The UCSC team is supported in part by NASA grants 14-WPS14-0048, NNG16PJ34C, NNG17PX03C, NSF grants AST-1518052 and AST-1815935, NASA through grant number AR-14296 from the Space Telescope Science Institute, which is operated by AURA, Inc., under NASA contract NAS 5-26555, the Gordon \& Betty Moore Foundation, the Heising-Simons Foundation, and by fellowships from the Alfred P. Sloan Foundation and the David and Lucile Packard Foundation to R.J.F. Some of the computations in this paper were run on the Odyssey cluster supported by the FAS Division of Science, Research Computing Group at Harvard University. The Pan-STARRS1 Surveys (PS1) and the PS1 public science archive have been made possible through contributions by the Institute for Astronomy, the University of Hawaii, the Pan-STARRS Project Office, the Max-Planck Society and its participating institutes, the Max Planck Institute for Astronomy, Heidelberg and the Max Planck Institute for Extraterrestrial Physics, Garching, The Johns Hopkins University, Durham University, the University of Edinburgh, the Queen's University Belfast, the Harvard-Smithsonian Center for Astrophysics, the Las Cumbres Observatory Global Telescope Network Incorporated, the National Central University of Taiwan, the Space Telescope Science Institute, the National Aeronautics and Space Administration under Grant No. NNX08AR22G issued through the Planetary Science Division of the NASA Science Mission Directorate, the National Science Foundation Grant No. AST-1238877, the University of Maryland, Eotvos Lorand University (ELTE), the Los Alamos National Laboratory, and the Gordon and Betty Moore Foundation.

\bibliography{mybib.bib}

\end{document}